# Spatiotemporal relative risk distribution of porcine reproductive and respiratory syndrome virus in the United States

**Felipe Sanchez[1,2], Jason A. Galvis[1], Nicolas Cardenas[1], Cesar Corzo[3], Christopher Jones[2], Gustavo Machado[1,2]\***

[1]Department of Population Health and Pathobiology, College of Veterinary Medicine, North Carolina State University, Raleigh, NC, USA.

[2]Center for Geospatial Analytics, North Carolina State University, Raleigh, NC, USA

[3]Veterinary Population Medicine Department, College of Veterinary Medicine, University of Minnesota, St Paul, MN, USA.

**\* Correspondence: Gustavo Machado, gmachad@ncsu.edu**

## Abstract

Porcine reproductive and respiratory syndrome virus (PRRSV) remains widely distributed across the U.S. swine industry. Between-farm movements of animals and transportation vehicles, along with local transmission are the primary routes by which PRRS is spread. Given the farm-to-farm proximity in high pig production areas, local transmission is an important pathway in the spread of PRRSV; however, there is limited understanding of the role local transmission plays in the dissemination of PRRSV, specifically, the distance at which there is increased risk for transmission from infected to susceptible farms. We used a spatial and spatiotemporal kernel density approach to estimate PRRSV relative risk and utilized a Bayesian spatiotemporal hierarchical model to assess the effects of environmental variables, between-farm movement data and on-farm biosecurity features on PRRSV outbreaks. The maximum spatial distance calculated through the kernel density approach was 15.3 km in 2018, 17.6 km in 2019, and 18 km in 2020. Spatiotemporal analysis revealed greater variability throughout the study period, with significant differences between the different farm types. We found that downstream farms (i.e., finisher and nursery farms) were located in areas of significant-high relative risk of PRRS. Factors associated with PRRSV outbreaks were farms with higher number of access points to barns, higher numbers of outgoing movements of pigs, and higher number of days where temperatures were between 4°C and 10°C. Results obtained from this study may be used to guide the reinforcement of biosecurity and surveillance strategies to farms and areas within the distance threshold of PRRSV positive farms.

## 1    Introduction

Porcine reproductive and respiratory syndrome virus (PRRSV) remains widely distributed across the U.S. swine industry (Galvis, Corzo, Prada et al., 2021; Jara et al., 2021; Sanhueza et al., 2020; Perez et al., 2019). Disease surveillance, vaccination strategies, and biosecurity protocols have played a key role in curving PRRSV outbreaks; however, variants of the endemic North American (NA-type, type 2) and the European (EU-type, type 1) strain periodically cause outbreaks that lead to significant economic losses (Kikuti et al., 2021; Jiang et al., 2020; Valdes-Donoso et al., 2018; van Geelen et al., 2018; Holtkamp et al., 2013; Benfield et al., 1992). Outbreaks of PRRSV in the U.S. have been shown to exhibit seasonal patterns throughout the country, but vary among swine-producing regions (Jara et al., 2021; Sanhueza et al., 2020; Alkhamis et al., 2018; Sanhueza et al., 2019; Arruda, Sanhueza et al., 2018). In the southeastern U.S., PRRSV outbreak patterns are typically characterized by a unimodal peak occurring during the fall and winter months, followed by decreases in incidence during the late spring and summer months (Jara et al., 2021; Sanhueza et al., 2020; Alkhamis et al., 2018; Arruda, Vilalta et al., 2018; Tousignant et al., 2015). Summer outbreaks, while less common, occur sporadically and vary by year (Kikuti et al., 2021; Sanhueza et al., 2020).

The spread of PRRSV is predominantly governed by direct contacts facilitated by the movement of infected pigs between farms, and indirect contacts also referred to as local transmission or area spread, which encompasses several unmeasured between-farm dynamics occurring at close geographical proximity (Galvis et al., 2022; Galvis, Corzo and Machado, 2021; Galvis, Corzo, Prada et al., 2021; Jara et al., 2021; Machado et al., 2020; VanderWaal et al., 2020; Dee et al., 2020; Otake et al., 2010; Dee et al., 2003). Despite local transmission being the least understood transmission pathway of many infectious diseases in humans and animals (Benincà et al., 2020), several epidemiological processes have been attributed to contributing to the local transmission of PRRSV including, the between-farm movement of contaminated personnel (Galli et al., 2022; Pitkin et al., 2009), trucks delivering pigs and feed (Galvis et al., 2022; Galli et al., 2022; Ruston, 2021), animal by-products delivered via feed (Galvis et al., 2022; Niederwerder, 2021; Dee et al., 2020; Ochoa et al., 2018), equipment (e.g., boots, coveralls, bleeding equipment) (Pitkin et al., 2009), and airborne viral particle dispersal (Kanankege et al., 2022; Li et al., 2021; Arruda et al., 2019; Otake et al., 2010; Dee et al., 2009; Otake et al., 2002). However, differentiating the contribution of each process remains highly challenging. Moreover, the distance at which each process poses a greater risk to neighboring farms remains poorly understood but is fundamental to the understanding of between-farm transmission dynamics (Shea et al., 2020). Between-farm transmission mechanisms acting on a local scale may vary in relation to the distance between farms and have been reported to range from one km to 35 km (Kanankege et al., 2022; Galvis et al., 2022; Galvis, Corzo, Prada et al., 2021; Jara et al., 2021; Machado et al., 2020; Arruda, Sanhueza et al., 2018; Lee et al., 2017; Pileri and Mateu, 2016; Dee et al., 2009; Otake et al., 2002; Christianson and Joo, 1994). Some of the local transmission mechanisms are also influenced by local environmental conditions (e.g., temperature, relative humidity, pH), genetic diversity of PRRSV, differences in management and biosecurity levels at different farm types, pig density, and the spatial proximity of susceptible farms to infected farms (farm density) (Pileri and Mateu, 2016; Jacobs et al., 2010; Dee et al., 2003, 2002). Given the high density of farms and pigs in intensive pig production areas across the U.S., a better understanding of the distance at which the risk of PRRSV transmission from infected to susceptible farms is increased may support and inform the implementation of targeted biosecurity enhancement and surveillance strategies (Moeller et al., 2022; Lambert et al., 2012).

In this study, we use an adaptive kernel density approach to derive spatial and spatiotemporal estimates of the variation in PRRSV relative risk. Using the kernel density estimate approach, we 1) define the maximum spatial distance at which farms may spread PRRSV based on the proximity of susceptible farms to infected farms and 2) identify farm types with elevated risk for local transmission of PRRSV. Secondly, we implemented a Bayesian spatiotemporal hierarchical model to account for environmental, on-farm biosecurity features, pig density, farm density, and between-farm contact networks metrics to 3) identify factors associated with the risk of PRRSV local transmission.

## 2    Materials and Methods

## 2.1 PRRSV data source and processing

PRRSV outbreak data for all production types used in this study were obtained from the Morrison Swine Health Monitoring Project (MSHMP) (Perez et al., 2019). Outbreak data collection was performed by each production company during outbreak investigations or routine surveillance activities and shared with MSHMP (Perez et al., 2019). Data obtained includes information on farm-level outbreaks between November 1st, 2017, through December 31st, 2020, from 2,293 farms belonging to three non-commercially related pig production companies in a dense pig production region of the U.S. Information about each farm includes, pig capacity, a unique farm identification number, geographical coordinates (hereafter geolocations), production type, and date of confirmed PRRSV outbreak via PRRSV positive laboratory results. Additionally, the distance between farms was calculated using farm geolocations. Production types in our farm population (n = 2,293) were classified as finisher (n = 1,458 and includes wean-to-finish, and feeder-to-finish), nursery (n = 468), isolation (n = 33), boar stud (n = 15), and sow (n = 319 and includes farrow, farrow-to-wean, and farrow-to-feeder farms).

Farms were divided into cases and controls, where cases were defined as any farm that reported an outbreak during the time period of interest, and controls are farms that did not report an outbreak. PRRSV case and control data were split into years (2018, 2019, and 2020) and a seasonal classification (PRRSV season). We defined the PRRSV season as a six-month period from November 1st through May 31st, which represents a time period where increases in farm-level PRRSV incidence have been previously described for the region of the U.S. considered in this study (Sanhueza et al., 2020; Tousignant et al., 2015).

## 2.2 Spatial PRRSV relative risk

Spatial kernel density-ratio, also known as spatial "relative risk" (hereafter risk), is an exploratory approach used to describe the geographical variation in disease risk based on the distribution of PRRSV outbreaks (cases) and the underlying at-risk (controls) population (Davies et al., 2018; Davies and Hazelton, 2010; Kelsall and Diggle, 1995; Bithell, 1990). PRRSV risk was estimated for each farm location ($x = \{x1, \ldots, x_n, n = 2,293 \: farms\}$) in each year and PRRSV season. Farms can report several PRRSV outbreaks in a given year or PRRSV season; however, for the spatial risk analysis, we defined cases as farms that reported at least one PRRSV outbreak, and controls as the remaining farms that did not report an outbreak for a given year and PRRSV season (Davies et al., 2018). We identified a total of 245 cases in 2018, 190 cases in 2019, and 165 cases in 2020. For the PRRSV seasons, a total of 227 cases in the 2017 - 2018 PRRSV season, 167 cases in the 2018 - 2019 PRRSV season, and 148 cases in the 2019 - 2020 PRRSV season were used. A nonparametric kernel density-ratio approach was used to estimate the risk $\widehat{p}(x)$ for each farm location ($x$) in each year and PRRSV season as follows:

$$\widehat{p}(x) = log \: \hat{f}(x) - log \: \hat{g}(x)$$

where $\hat{f}(x)$ represents the log density estimates of cases and $\hat{g}(x)$ represents the log density estimates of controls. The natural log is used to symmetrize the treatment of the density estimate ratios, with $\widehat{p}(x) > 0$, representing areas of medium to high PRRSV risk (high concentrations of cases relative to controls), and $\widehat{p}(x) < 0$, representing areas of low PRRSV risk (low concentration of cases relative to controls) (Davies et al., 2018; Fernando and Hazelton, 2014; Davies and Hazelton, 2010; Kelsall and Diggle, 1995). Calculating spatial risk relies on the selection of an optimal bandwidth (the maximum distance at which local transmission is unlikely to occur) which directly drives the decline of the risk probability (kernel) given the geolocation of a farm (Davies et al., 2018, 2016; Davies and Hazelton, 2010). Given the heterogeneous distribution of farm density in our study area, we used an adaptive smoothing approach that allows the bandwidth of each kernel to vary depending on the density of farms (cases and controls) at a given farm geolocation (Abramson, 1982). This method reduces smoothing at locations where the density of farms is high (e.g., 10 - 20 farms per 5 km²), and increases the amount of smoothing in areas where farm density is low (e.g., 1 - 5 farms per 5 km²) (Abramson, 1982). Adaptive smoothing requires the selection of pilot and global bandwidths, where the pilot bandwidths (i.e., cases and controls have a separate fixed distance), and the global bandwidth (i.e., cases and controls have a shared fixed distance), which is a smoothing parameter that adjusts the pilot bandwidth in areas where case and control densities are similarly distributed (Davies and Hazelton, 2010). Here, we compared two different approaches—asymmetric and symmetric adaptive smoothing—for the selection of the pilot bandwidths (Supplementary Material S1 and Supplementary Material Figure S1 - S6). Pilot and global bandwidths were then used to calculate $\hat{f}(x)$ and $\hat{g}(x)$. Spatial risk (Equation 1) was then calculated by using $\hat{f}(x)$ and $\hat{g}(x)$, and applying a uniform edge-correction, which accounts for the probability loss of farm geolocations close to the boundary of the study region (Davies and Hazelton, 2010; Davies et al., 2018). Lastly, we used 1,000 iterations of Monte Carlo simulations to delineate areas of significant spatial risk ($p < 0.05$) (Hazelton and Davies, 2009). Farms within areas of significant spatial risk were extracted and summarized as the count of case or control farms falling within areas of significant spatial risk by farm type.

## 2.3    Spatiotemporal PRRSV relative risk

The spatiotemporal risk of PRRSV was estimated in weekly time steps of cases for each year and PRRSV season, thus cases were defined as farms with at least one outbreak per week and controls as farms that did not report outbreaks for a given week. The entire farm population (n = 2,293) is considered in each weekly time step. A total of 438 cases with an average of 8.76 cases/week were used in 2018, 279 cases with an average of 5.47 cases/week in 2019, and 238 cases with an average of 4.67 cases/week in 2020. Similarly, a total of 382 cases with an average of 12.7 cases/week were used for the 2017 - 2018 PRRSV season, 231 with an average of 7.45 cases/week in the 2018 - 2019 PRRSV season, and 190 with an average of 6.33 cases/week in the 2019 - 2020 PRRSV season. In contrast to the spatial risk, the spatiotemporal risk uses spatial and temporal bandwidths derived from farm geolocations of cases to generate density estimates of cases, while density estimates for controls were generated using only the spatial bandwidth previously calculated for cases since we assume the farm population to be static between November 1st, 2017 and December 31st, 2020 (Davies and Lawson, 2019; Fernando and Hazelton, 2014; Terrell, 1990). Thus, conditional spatiotemporal risk surfaces were derived as:

$$\hat{\rho}(x|t) = log\ \hat{f}(x,t) - log\ \hat{f}(t) - log\ \hat{g}(x)$$

where $\hat{\rho}(x|t)$ is the conditional risk, $f(x,t)$ is the log density estimates of cases at a given time step $t = (t1,\ldots,t_n, n = 52\ weeks\ per\ year;\ n = 30\ weeks\ per\ PRRSV\ season)$, $f(t)$ is an estimator for the marginal temporal case density, and $g(x)$ is the static spatial log density of the controls (Fernando and Hazelton, 2014). One thousand iterations of Monte Carlo simulations were used to delineate areas of significant spatiotemporal risk (p < 0.05) (Hazelton and Davies, 2009).

Spatiotemporal risk values generated for the entire farm population at each time step ($t$) were extracted to individual farm geolocations ($x$). Farms were then classified as low, medium, and high risk based on quantiles of the spatiotemporal risk distribution of all the farms in each year and PRRSV season. Spatiotemporal risk above 60% of the risk distribution was considered as the exceedance risk threshold (Richardson et al., 2004) since it represented a midpoint between lower (negative) and higher (positive) risk values for the years and PRRSV season. Thus, farms with risk below 60% of the risk distribution were categorized as low risk, 61% to 80% quantile as medium risk, and 81% to 100% quantile as high risk for each year and PRRSV season (Supplementary Material Figures S7 - S12).

## 2.4    Priority Index

The priority index (PI) is a metric that has been used to facilitate the communication of spatiotemporal risk (Baquero and Machado, 2018). The aim of the PI is to provide an easily interpretable metric, represented as an ordered percentage that indicates the level of prioritization that should be given to a farm based on the estimated risk. The priority index was calculated from the spatiotemporal risk weekly estimates as:

$$PI = \hat{\rho}(x|t) \,/\, max(\,\hat{\rho}(x|t)\,) * 100$$

where the PI of a farm is a percentage based on the risk $\hat{\rho}(x|t)$ of a farm in reference to the maximum risk value of the farm population. Priority indices calculated for each farm were further reclassified as low (0 - 30%), medium (31% - 60%), and high (61% -100%) priority classifications based on quantiles. Priority index classifications were then summarized by farm types for each year and PRRSV season time periods and used to set the priority risk order of each farm type.

## 2.5    Bayesian spatiotemporal hierarchical model framework

We fit a Bayesian spatiotemporal hierarchical model of PRRSV weekly outbreak data to account for three on-farm biosecurity features, six between-farm contact network metrics, six environmental variables, farm density, and pig density (Figure 1 and Supplementary Material Table S1). A total of 2,072 farms out of our 2,293 farms were used in the Bayesian spatiotemporal hierarchical model, since 217 farms lacked between-farm contact data and four lacked environmental data. Additionally, between-farm contact data was not available for the entire study period; therefore, the Bayesian spatiotemporal hierarchical model was implemented for the year 2020. Farm geolocations $i$ ($i = 1,...n, n = 2,072\ farms\ in\ the\ year$ 2020) were defined as $Y_i$ where cases of PRRSV are represented as $Y_i = 1$ and controls are $Y_i = 0$ for each week in the year 2020. The generalized hierarchical spatiotemporal model was implemented as a logistic regression, where $Yi$ follows a binomial distribution:

$$Y_i = Logit\ (\mu_i)$$

and linear predictors were constructed as:

$$Logit\ (\mu_i) = \boldsymbol{\alpha} + \beta_i \cdot X_i + \upsilon\ (s_i) + week\ (s_i) + \omega\ (s_i)$$

where $\mu$ represents the probability of a PRRSV outbreak, $\alpha$ the model intercept, $\beta_i \cdot X_i$ describes the matrix of covariates, $\upsilon\ (s_i)$ is an independent and identically distributed (iid) random effect to account for variation between individual farms, $week(s_i)$ is an *iid* random effect to account for variation between weeks, and $\omega\ (s_i)$ represents a spatial random field (Gaussian field) to account for spatial errors (Krainski et al., 2018).

Briefly, the regression analysis was implemented with a stochastic partial differential equation (SPDE) model using integrated nested Laplace approximations (INLA) (van Niekerk et al., 2021; Elias T. et al., 2019; Bakka et al., 2018; Lindgren et al., 2011; Rue et al., 2009). The process first requires the creation of a mesh of Delaunay triangulations, which includes the specification of the maximum triangle edge length, and the model domain boundary. The resulting mesh (Supplementary Material Figure S13) consisted of 4,774 triangle vertices, where the model domain boundary was defined by a polygon representative of our study area in which the maximum triangle edge length was specified as 10 km within the inner domain and 20 km in the outer domain (Krainski et al., 2018).

The INLA default priors were used; therefore, penalized complexity (PC) priors were used for the spatial random fields where the spatial range and standard deviation quantile and probability tailored to be higher than 1 is 0.01 (Fuglstad et al., 2019; Simpson et al., 2017; Rue et al., 2009). Model fixed effect outputs were exponentiated and presented as odds ratio (Santos Baquero, 2018; Blangiardo and Cameletti, 2015). The sensitivity of priors to the posterior random field values was examined by comparing the random posterior mean distribution values of PC priors [(1, 0.01), (0.32, 0.01)] against log-gamma priors [(1, 0.05), (1, 0.001)] (Supplementary Material Figure S14).

## 2.6    Bayesian spatiotemporal hierarchical model data preparation

Variables considered in our Bayesian spatiotemporal hierarchical model framework focus on local transmission mechanisms, and environmental or anthropogenically mediated factors that may contribute to increases or decreases in risk of PRRSV outbreaks (Figure 1 and Supplementary Material Table S1) (Galvis, Corzo, and Machado, 2021; Galvis, Corzo, Prada et al., 2021; Jara et al., 2021; Machado et al., 2020; VanderWaal et al., 2020; Sanhueza et al., 2020; Cutler et al., 2012; Satoshi Otake et al., 2010; Hermann et al., 2007). On-farm biosecurity feature data were extracted for each farm from a database of Secure Pork Supply (SPS) biosecurity plans assembled by the Rapid Access Biosecurity app (RABapp™) (Machado, 2023) and included: the count of site entries, count of perimeter buffer area access points (PBAAP), and count of lines of separation access points (LOSAP) (Supplementary Material Figure S16 and Supplementary Material Table S1). In addition to on-farm biosecurity features, we included pig capacity, and farm density, which was calculated by creating a spatial buffer of 17 km around each farm location and counting the number of farms within the buffer. A spatial buffer of 17 km was used based on findings from the spatiotemporal kernel density approach discussed in further detail in section 3.2.

A directed and static network was reconstructed from between-farm pig movement data between January 1st, 2020, and December 31st, 2020, and represented as a graph g = (V, E), where V represents the nodes (farm) of the network and E represents the contact between two nodes or edges of the network. The unique farm identification number in each origin and destination movement record was used to form the edges of the network (Brandes and Erlebach, 2005). Between-farm contact network metrics: in-degree, out-degree, PageRank, clustering coefficient, closeness centrality, and betweenness were calculated to characterize node and network-level features of the directed, static network and are described in Supplementary Material Table S1. A total of 217 farms were missing pig movement data in 2020, and thus were excluded from this dataset. Therefore, between-farm pig movement data belonging to 2,072 farms was used to calculate between-farm contact network metrics considered in the Bayesian spatiotemporal hierarchical model framework (Figure 1).

Individual farm geolocations were used to extract environmental variables: weekly enhanced vegetation index (EVI), downloaded from the National Aeronautics and Space Administration (NASA), Moderate Resolution Imaging Spectroradiometer (MODIS) Land Products (Busetto and Ranghetti, 2016), and yearly averages of aboveground biomass density (AGBD), canopy height, and land surface elevation, downloaded from the Oak Ridge National Laboratory, Distributed Active Archive Center for Biogeochemical Dynamics website (ORNL DAAC) (ORNL DAAC, 2022). These variables are meant to represent topographical or vegetative buffers around farms that reduce or facilitate the spread of airborne particulate matter and PRRS virus (Jara et al., 2021; Guo et al., 2019; Arruda et al., 2017; Adrizal et al., 2008; Tyndall and Colletti, 2006). Similarly, farm geolocations were used to extract daily average land surface temperature, and relative humidity data from Daymet: Daily Surface Weather Data (Thornton et al., 2020). Temperature and relative humidity have been shown to impact the infectivity and stability of PRRSV (Cutler et al., 2012; Jacobs et al., 2010; Hermann et al., 2007), and were included in our model as the count of days a farm geolocation $i$ was associated with temperatures between 4°C and 10 °C (T[4°C,10°C] ) (e.g., farm geolocation $i$ had 20 days of T[4°C,10°C] over the studied period), and the count of days a farm geolocation $i$ was associated with relative humidity below 20% (RH < 20%) (e.g., farm geolocation $i$ was 30 days on RH < 20% over the studied period). All variables listed above were downloaded at a 1 km x 1 km spatial resolution and are described in further detail in Supplementary Material Table S1. A total of 71 farm geolocations were outside the data availability range of the AGBD, canopy height, or land surface elevation data products; therefore, data from the nearest farm (average = 5 km [min = 446 m; max = 9.8 km]) within a 10 km radius with complete data were used. Four farms were beyond the 10 km threshold and were excluded from the analysis (Supplementary Material Figure S15).

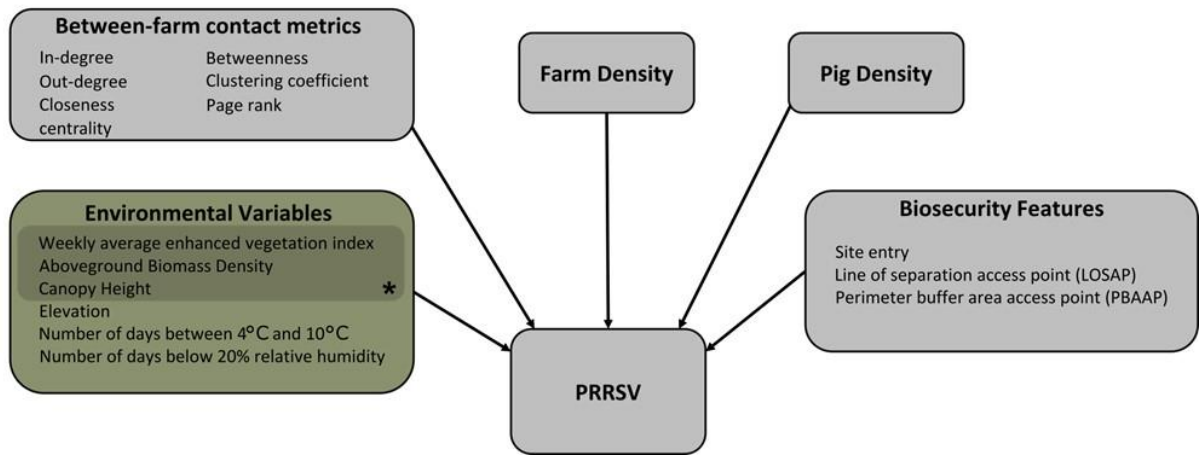

**Figure 1**. The conceptual model framework of the Bayesian spatiotemporal hierarchical model showing the directional relationship between variables and PRRSV outbreaks. * Variables representing vegetation buffers around farms.

## 2.7 Bayesian spatiotemporal hierarchical model variable selection and model comparison

All variables considered in our model framework (Figure 1) were first examined via univariate analysis following the model established in Equation 5, and significance was determined by the 95% credible intervals (CI) in which estimates did not cross one, and the model fit was evaluated using the Watanabe-Akaike information criteria (WAIC). Before selecting variables for the multivariate model, multicollinearity between variables was examined by calculating Pearson's correlation coefficient (r), where significant ($p < 0.05$) correlations above 0.6 were considered highly correlated and would limit our ability to determine individual effects on PRRSV outbreaks. The variables T[4°C,10°C] and RH < 20% (r = 0.98), biosecurity features LOSAP and PBAAP (r = 0.79), and network metrics, page rank and in-degree (r = 0.69) were highly correlated. All other variables were below 0.6 or had insignificant correlations. Among highly correlated variables, the variable yielding a lower WAIC in the univariate analysis was selected for the multivariate model variable selection process. A multivariate backward elimination process was carried out for all selected variables and the best-fitting model was selected based on the WAIC. Given the high density of farms in our study area, farm density was included in the multivariate analysis as a confounding factor.

All data extraction, processing, and analyses presented in this work were performed in the R (4.2.1) programming language (Core Team, 2021) using the packages: tidyverse (Wickham et al., 2019), sf (Pebesma, 2018), sp (Bivand et al., 2013), spatstat (Baddeley et al., 2016), sparr (Davies et al., 2011), raster (Hijmans et al., 2022), igraph (Csárdi et al., 2016), MODIStsp (Busetto and Ranghetti, 2016), daymetr (Hufkens et al., 2018), INLA (Bakka et al., 2018), inlabru (Bachl et al., 2019), and INLAoutputs (Santos Baquero, 2018).

## 3    Results

### 3.1    Spatial PRRSV relative risk

The median distance between farms reporting PRRSV outbreaks was 66.7 km (interquartile range (IQR): 39.4 km - 109.3 km) for 2018, 70 km (IQR: 40.4 km - 114 km) for 2019, 61.4 km (IQR: 36.6 km - 98 km) for 2020, and 66 km (IQR: 39 km - 106.4 km) for all years combined. For the PRRSV seasons, the distance between PRRSV cases was 64.6 km (IQR: 38 km - 107.5 km) for the 2017 - 2018 PRRSV season, 70.7 km (IQR: 40.2 km - 120.5 km) for the 2018 - 2019 PRRSV season, and 63.4 km (IQR: 37.3 km - 102 km) for the 2019 - 2020 PRRSV season. The maximum distance the spatial PRRSV risk extended to was, on average, 14.8 km for both the annual and PRRSV seasons. A total of 377 farms in 2018, 91 in 2019, and 321 in 2020 were within high risk areas ($p < 0.05$) (Table 1). Among the different farm types, sow farms consistently had a higher number of infected farms within areas of significant high risk in both the annual and PRRSV season time periods, while finisher and nursery farms had more control farms (Table 1 and Supplementary Material Table S2, S3, and S4). Comparison among years, 2018 (n = 85) had the greatest number of PRRSV infected farms within significant high risk areas compared to 2019 (n = 21) and 2020 (n = 57) (Table 1).

**Table 1.** Yearly count of cases and controls by farm type within significant ($p < 0.05$) high risk areas estimated using a spatial asymmetric adaptive smoothing approach for 2,293 farms (n = number of farms per farm type) in a dense pig production region of the U.S.

| Year | Sow (n = 319) | | Nursery (n = 468) | | Finisher (n = 1,458) | | Isolation (n = 33) | | Boar Stud (n = 15) | |
|---|---|---|---|---|---|---|---|---|---|---|
| | Cases | Controls | Cases | Controls | Cases | Controls | Cases | Controls | Cases | Controls |
| **2018** | 34/319 | 49/319 | 13/468 | 79/468 | 21/1458 | 126/1458 | 0 | 2/33 | 0 | 4/15 |

| 2019 | 27/319 | 21/319 | 6/468 | 52/468 | 5/1458 | 85/1458 | 0 | 1/33 | 0 | 2/15 |
| 2020 | 44/319 | 43/319 | 7/468 | 81/468 | 6/1458 | 135/1458 | 0 | 3/33 | 0 | 3/15 |

## 3.2    Spatiotemporal PRRSV relative risk

The spatiotemporal analysis revealed that the maximum distance PRRSV risk extended to was 15.3 km in 2018, 17.6 km in 2019, and 18 km in 2020, and for the PRRSV seasons, 13.6 km in the PRRSV 2017 - 2018, 19.2 km in the PRRSV 2018 - 2019, and 18.9 km in the PRRSV 2019 - 2020. Spatiotemporal risk estimates for the entire farm population in each time step were classified as high, medium, and low risk based on a 60% exceedance threshold where, on average, 20% of the farms were classified as high risk, 20.1% as medium risk, and 59.9% as low risk for all farm types and years combined (Supplementary Material Figures S7 - S12). Among farm types, boar stud farms were on average more frequently classified to be in high risk areas (30% IQR: 27% -38%) for the entire study period, followed by sow (29% IQR: 27% - 34%), nursery (19% IQR: 18% - 22%), finisher (14% IQR: 13% - 19%) and isolation farms (12% IQR: 9% - 24%) (Table 2). However, it is important to note that the higher percentages seen for boar stud farms may be driven by the fewer number of boar stud farms (n= 15) in the farm population. Spatiotemporal risk estimates for the PRRSV seasons revealed sow farms we more frequently classified to be in high risk areas (29% IQR: 24% - 36%) during this time period, followed by boar studs (27% IQR: 20% - 33%), nursery (21% IQR: 18% - 24%), isolation farms (12% IQR: 12% - 30%), and finisher (18% IQR: 16% - 25%) (Supplementary Material Table S5). Among all farm types, finisher farms (63% IQR: 54% - 74%) and nursery farms (57% IQR: 52% - 67%) were consistently classified to be in areas of low risk for both the yearly (Table 2) and PRRSV season time periods (Supplementary Material Table S5).

Our spatiotemporal analysis revealed a seasonal signal, marked by an increase in farms classified as being in high and medium PRRSV risk areas during the fall, winter, and spring months, with varying intensity between farm types, and years 2018 to 2019, and 2019 to 2020 (Figure 2 and Supplementary Material Figure S8). The signal onset of the seasonal pattern appears to begin at an earlier date (mid-summer [Week 28] to early fall [Week 35]) for the year 2019; however, it is not consistent among all farm types. Sow farms displayed an interesting pattern among farm types, with increases in risk during summer months (Week 20 - 35) (Figure 2 and Supplementary Material Figure S17). Nursery, finisher, boar stud, and isolation farms appear to show a similar summer increase in the year 2020, but it is not present for other years. Among all farm types, sow, finisher and nursery farms appear to more closely resemble each other in terms of seasonal risk. Boar stud and isolation show more erratic changes in risk, however, the large shifts in risk levels are related to the small number of farms for these farm types.

Results obtained from calculating the PI, which may be used to order farms in risk priority, revealed that 79.4% of the farms in 2018 were in the low PI category, 16.1 % were in the medium PI, and 4.5% were in the high PI category (Supplementary Material Table S6). Similarly, 63.7% in 2019 and 67.9% in 2020 were in the low PI category, 28% in 2019 and 23.9% in 2020 were in the medium PI category, and 8.35% in 2019 and 8.28% in 2020 were in the high PI category. Among the different farm types, sow farms consistently had the most farms in the high and medium risk category except boar stud farms in 2019 and 2020; however, as noted before, there are fewer boar stud farms as compared to sow farms in the study population. A similar proportion of farms with high, medium, and low PI overall and by farm type were observed for the PRRSV seasons (Supplementary Material Table S7).

**Table 2**. Percent of high, medium, and low PRRSV risk levels (median and interquartile range (IQR)) based on weekly risk estimates obtained from the spatiotemporal analysis by farm type for each year and for the entire study period.

| Period | Sow | | | Nursery | | | Finisher | | | Isolation | | | Boar Stud | | |
|---|---|---|---|---|---|---|---|---|---|---|---|---|---|---|---|
| | High | Med. | Low | High | Med. | Low | High | Med. | Low | High | Med. | Low | High | Med. | Low |
| **2018** | 24% (24%-29%) | 22% (17%-31%) | 51% (40%-57%) | 18% (16%-21%) | 21% (15%-30%) | 56% (50%-73%) | 17% (13%-25%) | 20% (10%-28%) | 59% (49%-77%) | 12% (9%-19%) | 39% (19%-45%) | 45% (39%-75%) | 27% (20%-27%) | 27% (20%-27%) | 53% (47%-60%) |
| **2019** | 33% (29%-37%) | 16% (14%-21%) | 49% (45%-53%) | 18% (17%-20%) | 16% (11%-23%) | 65% (55%-71%) | 14% (14%-20%) | 17% (6%-21%) | 71% (55%-79%) | 12% (12%-48%) | 21% (12%-33%) | 33% (27%-88%) | 40% (33%-40%) | 13% (7%-20%) | 47% (47%-47%) |
| **2020** | 30% (28%-35%) | 27% (16%-29%) | 44% (42%-48%) | 20% (19%-25%) | 22% (20%-28%) | 53% (51%-59%) | 15% (13%-16%) | 25% (21%-28%) | 61% (58%-64%) | 18% (6%-29%) | 42% (24%-48%) | 39% (33%-73%) | 27% (20%-33%) | 20% (13%-27%) | 53% (40%-60%) |
| **2018 - 2020 \*** | 29% (27%-34%) | 21% (14%-28%) | 48% (43%-53%) | 19% (18%-22%) | 21% (15%-26%) | 57% (52%-67%) | 14% (13%-19%) | 21% (13%-26%) | 63% (54%-74%) | 12% (9%-24%) | 35% (18%-45%) | 42% (33%-82%) | 30% (27%-38%) | 20% (13%-27%) | 47% (47%-53%) |

\* Median and IQR for all years combined.

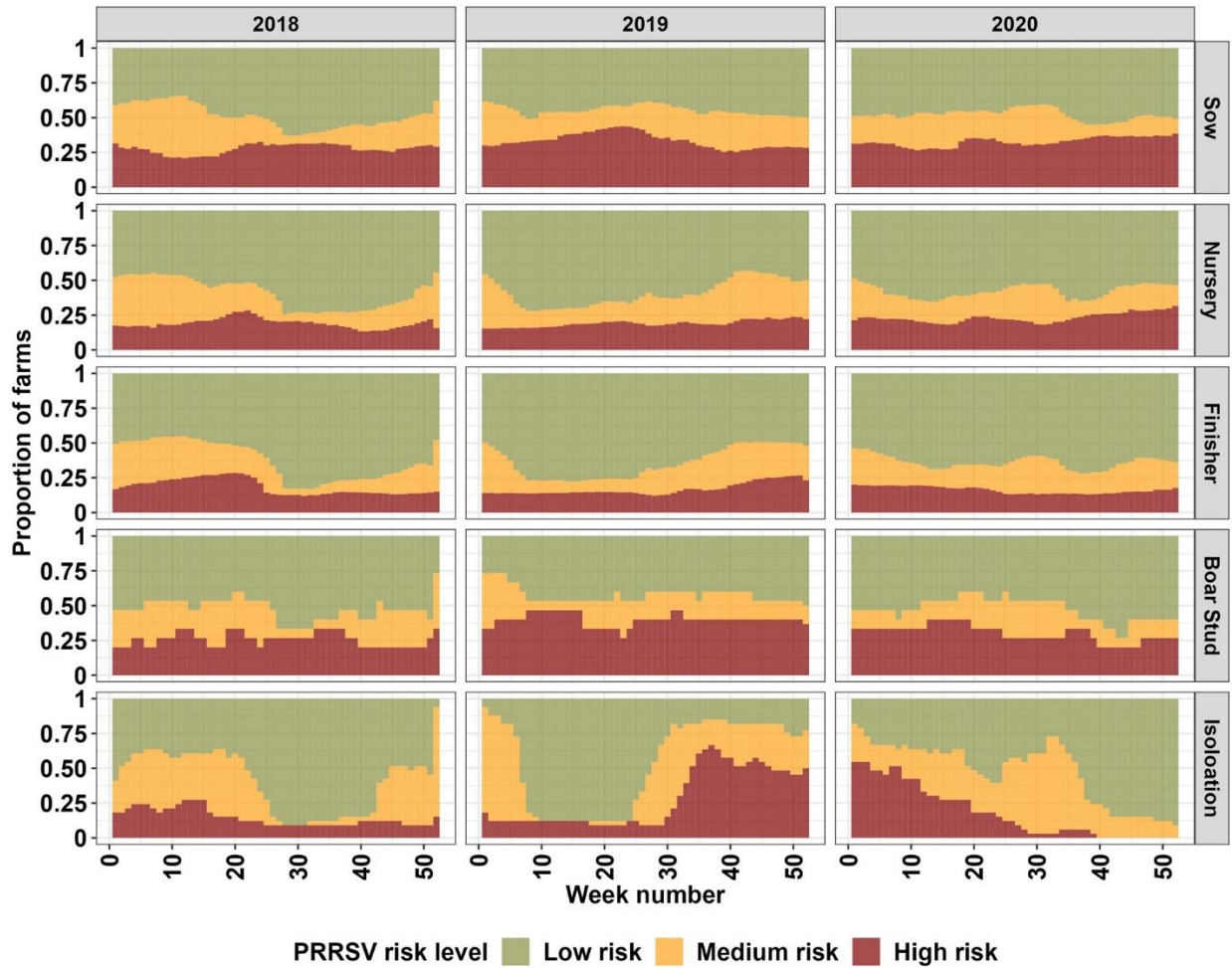

**Figure 2**. Farm type breakdown of high, medium, and low PRRSV risk levels for the entire farm population (2,293) based on a 60% exceedance relative risk threshold for each week (from 1 to 52 weeks) in the years 2018 through 2020.

### 3.3 Bayesian spatiotemporal hierarchical model

Results from the univariate analysis revealed that the animal movement network metric, out-degree, the number of LOSAP, number of days the temperature was between T [4°C,10°C], and relative humidity < 20% were significantly associated with PRRSV outbreaks (Table 3). The best fitting multivariate model (WAIC = 2,620) obtained through backward variable selection included the following variables: out-degree, LOSAP, T [4°C, 10°C], and farm density (Table 3). The strongest association was related to out-going movements, which resulted in an increase in the odds of PRRSV outbreaks by 1.11 times (Table 3). The second most associated variable was LOSAP, with an increase of 1.04 times the odds of PRRSV outbreaks for every additional LOSAP. Lastly, T [4°C, 10°C] resulted in a 1.01 increase in odds for every one unit increase in T [4°C, 10°C].

**Table 3.** Summary statistics of the fixed effects of the Bayesian spatiotemporal hierarchical models, showing odds ratio (OR), 25% - 97.5% credible intervals, and WAIC.

| Univariate | Multivariate |
|---|---|

| Covariate | CI 0.025 | OR | CI 0.975 | WAIC | CI 0.025 | OR | CI 0.975 | WAIC |
|---|---|---|---|---|---|---|---|---|
| Pig capacity | 1 | 1 | 1 | 2,704.40 | - | - | - | |
| Farm density | 0.99 | 1 | 1 | 2,703.80 | 1 | 1 | 1.01 | |
| EVI | 1 | 1 | 1 | 2,704.05 | - | - | - | |
| **T[4°C,10°C]** | **1.01** | **1.02** | **1.03** | **2,672.23** | **1.01** | **1.01** | **1.02** | |
| **RH < 20%** | **1.01** | **1.01** | **1.02** | **2,676.66** | | | | |
| Aboveground biomass density | 0.97 | 1 | 1.03 | 2,705.47 | - | - | - | |
| Canopy height | 0.69 | 0.96 | 1.30 | 2,709.50 | - | - | - | |
| Elevation | 0.99 | 1 | 1.01 | 2,701.34 | - | - | - | |
| In-degree | 0.95 | 1 | 1.01 | 2,706.15 | - | - | - | |
| **Out-degree** | **1.08** | **1.11** | **1.14** | **2,641.94** | **1.08** | **1.11** | **1.14** | |
| Closeness centrality | -0.14 | 37.44 | 229.68 | 2,704.82 | - | - | - | |
| Betweenness | 1 | 1 | 1 | 2,701.65 | - | - | - | |
| Clustering coefficient | 0.07 | 0.239 | 1.11 | 2,701.22 | - | - | - | |
| Page rank | 0 | 0 | 0 | 2,707.16 | - | - | - | |
| Site entry | $7.698183 \times 10^{14}$ | $4.102065 \times 10^{4}$ | $12.9776010^{10}$ | 13,017.82 | - | - | - | |
| PBAAP | 0.98 | 1.03 | 1.07 | 2,704.70 | - | - | - | |
| **LOSAP** | **1.02** | **1.05** | **1.08** | **2,695.96** | **1.00** | **1.04** | **1.07** | **2,620.15** |

## 4    Discussion

We estimated the maximum distance at which the risk of PRRSV is significantly high given the spatial proximity of farms reporting PRRSV outbreaks. We demonstrated through our spatial and spatiotemporal analysis that farms within an 11.9 km to 17 km radius of PRRSV positive farms were at greater risk of being infected due to proximity. PRRSV risk was higher during fall winter and early spring months with variation among the different farm types and years (Figure 2), which is consistent with seasonal patterns previously described for this region of the U.S. (Jara et al., 2021; Sanhueza et al., 2020; Arruda, Sanhueza et al., 2018; Alkhamis et al., 2018; Tousignant et al., 2015). Spatiotemporal risk estimates revealed that approximately 29% of sow farms were consistently located in areas of high risk. We have also demonstrated that outgoing animal movements (out-degree), the number of barn access points (LOSAP), and the number of days where temperatures were between 4°C and 10°C (T[4°C,10°C]) were risk factors for PRRSV outbreaks (Table 3).

Given the temporal dynamics of PRRSV, and in comparison to the spatial analysis which is a time-static representation of farms reporting PRRSV outbreaks for an entire year and/or PRRSV season, weekly PRRSV outbreak reports were used in our spatiotemporal analysis. Our spatiotemporal analysis showed that the risk of PRRSV transmission from infected farms was significant up to 17 km compared to 11.9 km in our spatial analysis. We attribute the increase in distance calculated in our spatiotemporal analysis to the density of cases to controls considered in each weekly time step, since we expect that the farm density modulates the size of the bandwidth radius, thus increasing the radius distance to compensate for the low density of cases (Davies et al., 2018, 2016). We remark that the spatial distribution of cases and controls alone may not be sufficient to fully explain PRRSV risk dynamics; therefore, we consider that the maximum distance of 17 km calculated in our spatiotemporal analysis reflects a close representation of PRRSV spatial risk in our study region, since risk estimates consider temporal patterns associated with PRRRV (Sanhueza et al., 2020; Arruda, Sanhueza et al., 2018; Alkhamis et al., 2018).

Our spatiotemporal analysis showed increases in PRRSV risk during the fall, winter, and early spring months, which aligns with previous findings for the dense pig production region considered in this study (Sanhueza et al., 2020; Arruda, Sanhueza et al., 2018; Alkhamis et al., 2018; Tousignant et al., 2015). The seasonal effect was consistently detected throughout the study period, but varied in intensity between years and farm types. In contrast to the typical seasonal patterns previously reported for the dense pig production region considered in this study and continued to be observed in this study, an increase in risk during the summer (Week 20 - 35) months was detected in sow farms for all years in our study period, with nursery and finisher farms displaying a similar pattern for the years 2018 and 2020, and boar stud and isolation farms only in 2020 (Figure 2). Summer PRRSV outbreaks in breeding and finisher farms have been previously reported (Kikuti et al., 2021; Sanhueza et al., 2020); however, we show that given the spatial proximity of farms in dense pig production areas, the risk for a PRRSV outbreak may propagate to different farm types. This information supports previous findings and highlights the importance of considering transmission dynamics between farm types during months outside the typical PRRSV season to help farm managers and veterinarians plan for enhanced biosecurity and surveillance (Sanhueza et al., 2020, 2019; Perez et al., 2019; Arruda, Vilalta et al., 2018; Alkhamis et al., 2018).

Transmission dynamics of PRRSV involve two main transmission pathways: direct contacts mediated by the movement of infected pigs between farms (Galvis et al., 2022; Galvis, Corzo and Machado, 2021; Machado et al., 2020; VanderWaal et al., 2020), and indirect contacts referred to as local transmission (Kanankege et al., 2022; Galvis et al., 2022; Galvis, Corzo and Machado, 2021; Galvis, Corzo, Prada et al., 2021; Jara et al., 2021; Dee et al., 2020; Arruda, Sanhueza et al., 2018; Otake et al., 2010; Dee et al., 2003). Local transmission encompasses several mechanisms of spread and is typically used to explain processes that occur over short geographical distances and cannot be attributed to direct contacts caused by live animal shipments among farms (Galvis et al., 2022; Galli et al., 2022; Galvis, Corzo, Prada et al., 2021; Ruston, 2021; Dee et al., 2020; Pitkin et al., 2009; Arruda et al., 2019). Airborne transmission of PRRSV has been suggested as a possible source of local transmission, especially in areas of high farm density; however, evidence has been inconsistent (Arruda et al., 2019). Experimental studies conducted to examine the distance airborne between-farm transmission of PRRSV may occur showed that PRRSV was recovered at a distance of 9.2 km (Otake et al., 2010; Dee et al., 2009). In a recent study, an atmospheric dispersion model was used to determine that farms within a distance of 25 km distance from a PRRSV positive farm were at high PRRSV risk (Kanankege et al., 2022). Dispersion models, such as the one developed by (Kanankege et al., 2022) may be invaluable tools when conducting outbreak investigations; however, as noted by the authors, further refinement to include environmental factors (e.g., temperature and humidity) and seasonal differences may yield more accurate estimates. In our study, both the spatial (11.9 km - 14.8 km) and spatiotemporal (17 km) distances calculated are within the distances proposed in (Kanankege et al., 2022; Otake et al., 2010), and given the high density (e.g., 10 - 20 farms per 5 km²) of farms in our study area, the potential for airborne transmission occurring in our study area cannot be ruled out. However, given the additional mechanisms (e.g., movement of contaminated transportation vehicles, equipment, and personnel) involved in local transmission that have been shown to contribute to the between-farm transmission of PRRSV (Galvis et al., 2022; Thakur et al., 2016; Pitkin et al., 2009), the small number of samples recovered through airborne transmission (Otake et al., 2010; Dee et al., 2009), and consideration of mechanical (presence of air filtration systems) or environmental factors (e.g., temperature, humidity, vegetation and slope) that may impact the survivability or infectivity of PRRSV, we agree with previous conclusions that airborne transmission is an infrequent or unlikely event (Arruda et al., 2019).

Breeding farms have been the center of most swine disease transmission studies (Moeller et al., 2022; Sanhueza et al., 2020; Passafaro et al., 2020; Kinsley et al., 2019; Lee et al., 2017; Pileri and Mateu, 2016). In this study, we have shown that the number of sow farms in high-risk areas was larger than all other farm types. A potential explanation as to why sow farms were consistently categorized as high risk may be due to the higher level of systematic testing performed at sow farms as compared to other farm types (Galvis et al., 2022; Galvis, Corzo, Prada et al., 2021; Perez et al., 2019). Higher levels of surveillance in sow farms increase the detection rate of PRRSV outbreaks, consequently increasing the number of analyzed PRRSV outbreaks in our study. Conversely, the underdetection of downstream farms, which has been noted in previous studies as a limitation to understanding PRRSV transmission dynamics (Galvis et al., 2022; Galvis, Corzo, Prada et al., 2021; Passafaro et al., 2020; Velasova et al., 2012; Lambert et al., 2012), is consistent with our findings in the spatial analysis where large numbers of PRRSV negative farms were within areas of significant high risk. A recent work investigated the association between PPRSV outbreaks and farm proximity to areas of high commingling (slaughterhouses) and found no association; however, the study only considered breeding herds, which highlights the need to consider other farm types that may contribute to disease circulation (Moeller et al., 2022). Our study showed both upstream and downstream farm types within areas of significant risk, and until systematic testing occurs in all farm types, estimations of PRRSV spatial risk will remain a challenge, especially for the estimations for growing pigs.

The most important risk factor associated with PRRSV outbreaks in this study was the movement of animals, which has been shown to be the dominant PRRSV transmission route (Black et al., 2022; Galvis et al., 2022; Galvis, Corzo and Machado, 2021; Passafaro et al., 2020; VanderWaal et al., 2020; Kinsley et al., 2019; Haredasht et al., 2017; Silva et al., 2019; Perez et al., 2015). Specifically, the effect was related to the out-going movements of animals, which is usually associated with farm types that have large and consistent outgoing shipments of pigs, such as sow and nursery farms. Such farms may usually take on the role of "movement hubs" in a network, thus facilitating the spread of diseases (Passafaro et al., 2020; Kinsley et al., 2019; Lee et al., 2017; Dorjee et al., 2013). The high number of out-going movements is supported in our findings, where the largest out-degree values were associated with infected nursery farms with a median value of 9 (IQR: 7 - 11), and 8 (IQR: 6 - 11) for controls (Supplementary Material S2 and Supplementary Material Table S8). Similarly, positive sow farms had out-degree median values of 6 (IQR: 4 - 10), and 6 (IQR: 3 - 10) for controls (Supplementary Material Table S8).

The second important variable was LOSAP, which can serve as potential entry pathway for the introduction of pathogens (Black et al., 2022; Lambert et al., 2012). Entry or exit through a LOSAP may involve several risk events such as garbage collection, equipment repair, and removal of cull sows that have been identified as relevant risk events associated with the introduction of diseases (Galli et al., 2022; Galvis et al., 2022, 2022; Pitkin et al., 2009; Silva et al., 2019, 2018). Among the different farm types, sow farms had the highest median LOSAP values with control farms having a median value of 5 (IQR: 1-16) LOSAP, and cases having a median value of 3 (IQR: 1-12) LOSAP.

Temperature and relative humidity have been shown to influence the survivability and optimal preservation of infectivity of PPRSV (Cutler et al., 2012; Hermann et al., 2007). Here, we showed that the increased number of days between 4°C and 10°C and the number of days a farm experienced relative humidity values below 20% increased the probability of PRRSV outbreaks. This is consistent with the seasonal signal of increased risk during the fall and winter months seen in our data and reported in previous research (Sanhueza et al., 2020; Tousignant et al., 2015). Lastly, we sought to expand on the investigation of the use of vegetation buffers as possible means to mitigate PRRSV transmission. EVI values between 41 and 45, which correspond to dense tree coverage that is consistent with evergreen broadleaf forests were shown to prevent the spread of PRRSV (Jara et al., 2021). We included EVI, AGB, and canopy height data to capture the structure of the vegetation around the farms; however, these variables were not significant. Given the coarse spatial resolution used in this study, and the benefits of using vegetation buffers to mitigate odor and pathogen emission and introduction (Jara et al., 2021; Guo et al., 2019; Adrizal et al., 2008), we remark that further exploration into the use of remotely sensed data to delineate vegetation buffers is warranted since the availability of imagery from satellites with high temporal and spatial resolution continues to become more accessible. Lastly, a previous study found the slope of the terrain to be associated with lower PRRSV incidence, with an elevation of 61 meters from sea level determined to be a safe range (Jara et al., 2021). We did not find a significant correlation between PRRSV outbreaks and land slope in our model; however, we remark that our results may in-part be related to the 1 km x 1 km spatial resolution of our data, and a finer spatial resolution should be explored.

## 4.1   Limitations and further remarks

The present study has limitations. Firstly, swine production is dynamic in nature, with farms being active and inactive throughout the years. During the time period considered in this study, 34/2,293 farms (1.5%) changed from active to inactive between Nov 1st, 2017 through December 31st 2020, thus we consider our results closely reflect the current status of the swine industry in our study region given the high level of participation of the swine production companies in our study. Another limitation in our study relates to how PRRSV surveillance systems differ between farm types, with sow farms usually conducting routine testing while downstream farm testing is not always performed systematically (Pileri and Mateu, 2016; Velasova et al., 2012; Lambert et al., 2012). Differences in systematic testing between the different farm types could have affected our risk estimations; however, we believe that the PRRSV database captured by MSHMP is still the best alternative to the currently available PRRSV datasets (Perez et al., 2019). For our spatiotemporal analysis, we arbitrarily chose the exceedance risk threshold to be at 60% since it aligned with the interpretation described in (Davies et al., 2018), where $\hat{p}(x) > 0$ represent areas of higher risk, and $\hat{p}(x) < 0$ areas of low risk. In future studies, other cutoff values should be explored.

Even though we had to restrict our risk factor analysis to 2020 due to limitations of the between-farm movement data, our results would likely be similar for other years, given how animal movements are vertically integrated in the U.S. (Sellman et al., 2022; Galvis et al., 2022; Galvis, Corzo, Prada et al., 2021; Passafaro et al., 2020; Machado et al., 2020). Environmental factors that are known to vary through time were averaged for an entire year, which might dilute the temporal differences in environmental conditions that may influence PRRSV transmission dynamics (Sanhueza et al., 2020; Arruda, Vilalta et al., 2018; Tousignant et al., 2015). However, results obtained from this study provide the important groundwork for further exploration of temporal differences related to factors associated with PRRSV local spread. Despite the limitations present in this study, here, we address an important gap related to the spatial range associated with the risk of PRRSV local transmission and estimate the maximum distance to which farms may become exposed and or infected from nearby infected farms. Both results could potentially be used to inform swine producers within areas of elevated risk to consider enhancing surveillance, sampling and disease control strategies (Alarcón et al., 2021; Perez et al., 2019, 2015). In addition, information gathered from this study may be used to calibrate future PRRSV transmission models by considering the calculated spatial bandwidths as the maximum transmission distance (Galvis et al., 2022; Galvis, Corzo and Machado, 2021; Galvis, Jones, Prada et al., 2021; Machado et al., 2020).

## 4.2   Conclusion

The results of this study suggest that farms within a 17 km radius of farms reporting PRRSV outbreaks are at greater risk of infection. We demonstrated that PRRSV outbreaks remain mostly seasonal, with differences in risk intensity between farm types. Our analysis also captured sporadic summer increases in risk, with differences between years and farm types. We found that sow farms had the highest number of cases within areas of significant high risk and were collocated with at-risk finisher and nursery farms. These findings suggest that downstream farms (i.e. finisher farms) may play an important role in maintaining the circulation of PRRSV within the farm population, and support the need for systematic testing among the different farm types. Lastly, out-going movement of pigs, the number of access points and temperature were significant risk factors of PRRSV outbreaks. Ultimately, we provide insights into PRRSV risk dynamics among farm types and establish a maximum distance for the risk of PPRSV local transmission, which could be used to inform targeted surveillance and disease control strategies and calibrate future PRRSV transmission models.

## Acknowledgements

The authors would like to acknowledge participating companies and veterinarians and the Morrison Swine Health Monitoring Project funded by the Swine Health Information Center.

**Author's contributions**

FS, CJ, JG, and GM conceived the study. FS, GM, CJ, NC, and JG participated in the design of the study and model formulation. CC coordinated the disease data collection by the Morrison Swine Health Monitoring Program (MSHMP). FS and JG conducted data processing, cleaning, designed the model, and simulated scenarios with the assistance of CJ, NC and GM. FS, JG, NC, GM, and CJ wrote and edited the manuscript. All authors discussed the results and critically reviewed the manuscript.

**Conflict of interest**

The authors declare that the research was conducted in the absence of any commercial or financial relationships that could be construed as a potential conflict of interest.

**Ethical statement**

The authors confirm the ethical policies of the journal, as noted on the journal's author guidelines page. Since this work did not involve animal sampling nor questionnaire data collection by the researchers, there was no need for ethics permits.

**Data Availability Statement**

The data that support the findings of this study are not publicly available and are protected by confidential agreements, therefore, are not available.

**Funding**

This project was funded by the Center for Geospatial Analytics and the College of Veterinary Medicine at North Carolina State University. The Morrison Swine Health Monitoring Project is a Swine Health Information Center (SHIC) funded project.

*Supplementary Material*

# Spatiotemporal relative risk distribution of porcine reproductive and respiratory syndrome virus in the United States

**Felipe Sanchez**[1,2]**, Jason A. Galvis**[1]**, Nicolas Cardenas**[1]**, Cesar Corzo**[3]**, Chris Jones**[2]**, Gustavo Machado**[1,2,*]

**\* Correspondence:** Gustavo Machado, gmachad@ncsu.edu

## S1. Spatial relative risk - Symmetric and Asymmetric adaptive smoothing

Asymmetric adaptive smoothing is achieved by calculating separate pilot bandwidths for cases and controls using point patterns constructed from case and control farm geolocations, and calculating the global bandwidth from the entire farm population. The asymmetric approach uses different pilot bandwidths generated from disproportionate numbers of cases to controls, which has been shown to produce artifacts in the calculation of RR (Davies et al., 2016). We contrast this approach with a symmetric adaptive smoothing approach, in which the pilot and global bandwidths are determined from a point pattern constructed from combined case-control data to account for the high spatial heterogeneity of farm density in our study area (Davies and Lawson, 2019; Davies et al., 2018, 2016; Davies and Hazelton, 2010; Lawson and Zhou, 2005; Prince et al., 2001; Keeling et al., 2001).

Comparing symmetric adaptive smoothing and asymmetric adaptive smoothing, our results show that the maximum distance the spatial PRRSV RR extended to was, on average, 11.9 km for both the annual and PRRSV seasons using the symmetric adaptive smoothing, and 14.8 km for asymmetric adaptive smoothing. Both symmetric and asymmetric approaches identified similar areas of significant risk of PRRSV (Supplementary Material Figure S1 - S6), with slight variations noted for the year 2019 (Supplementary Material Figure S2), and PRRSV season 2018 - 2019 (Supplementary Material Figure S5). Areas of significant high risk generated by the symmetric approach for both the year 2019 and PRRSV season 2018 - 2019 showed several additional areas of significant high risk as compared to the asymmetric approach. However, these areas were very small (0.69 km²); therefore, should be interpreted with caution (Davies et al., 2018, 2016) given that these areas correspond to areas of low farm density (e.g., 1 - 5 farms per 5 km²). While the asymmetric approach has been shown to produce artifacts in the calculation of RR as compared to the symmetric approach (Davies et al., 2016), we consider the asymmetric approach to be more reasonable since it is less impacted by the uneven spatial distribution of cases to controls (Davies et al., 2018, 2016).

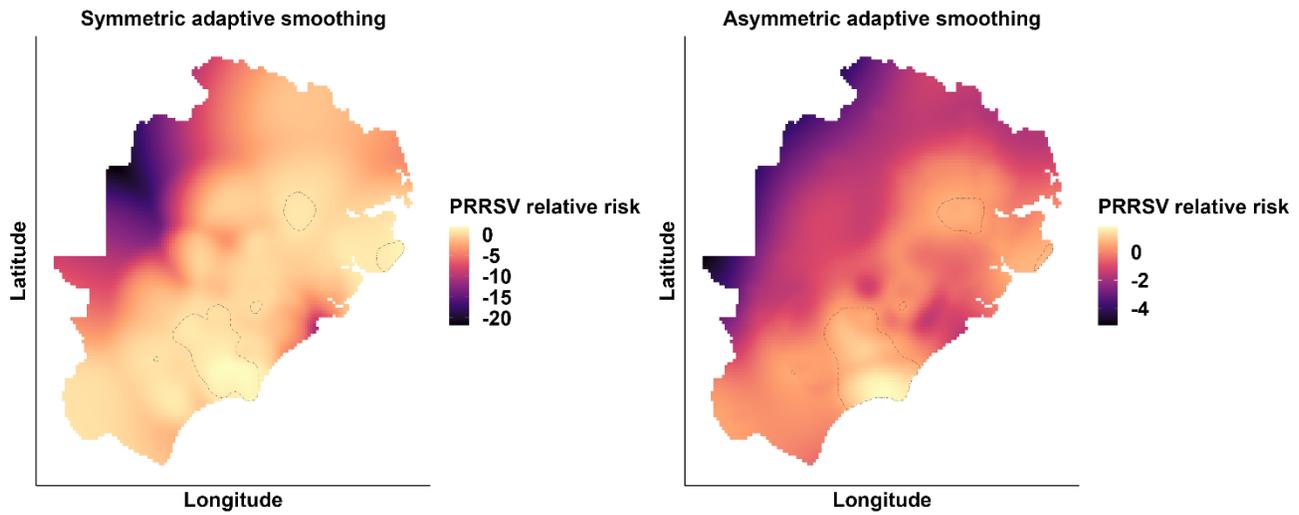

**Supplementary Material Figure S1**. Spatial relative risk estimates for 2018 using symmetric adaptive smoothing and asymmetric adaptive smoothing with 0.05 high risk tolerance contours denoted by a grey dash dot line.

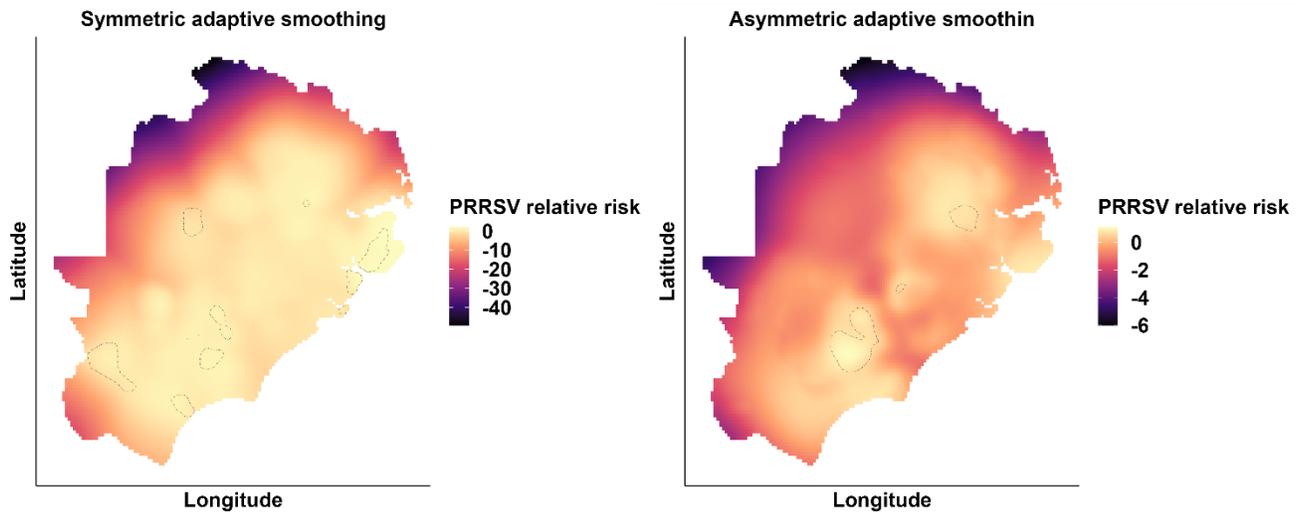

**Supplementary Material Figure S2**. Spatial relative risk estimates for 2019 using symmetric adaptive smoothing and asymmetric adaptive smoothing with 0.05 high risk tolerance contours denoted by a grey dash dot line.

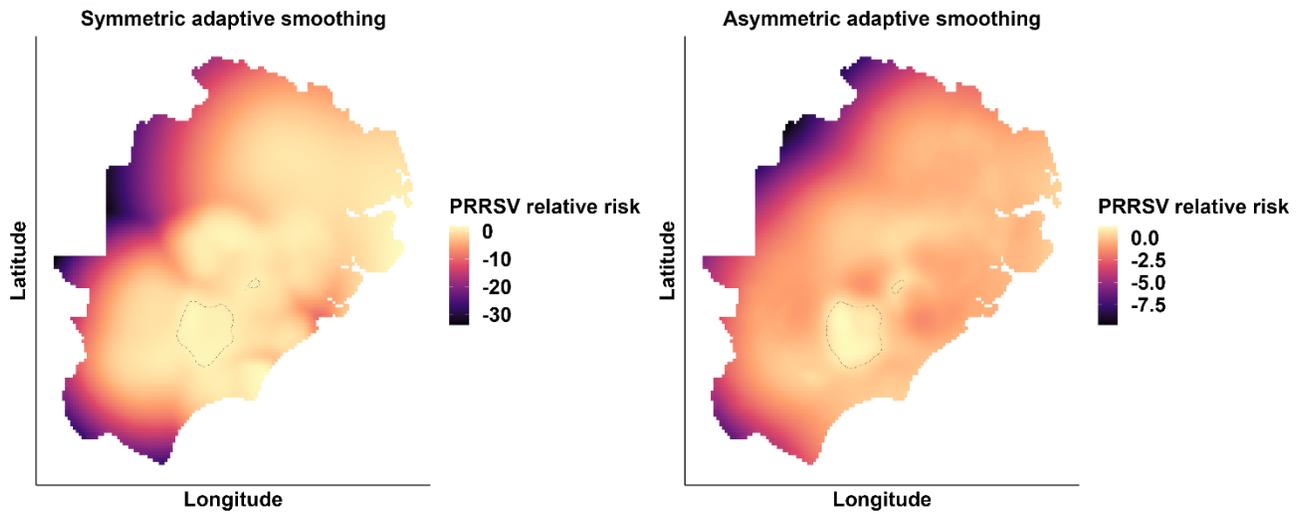

**Supplementary Material Figure S3**. Spatial relative risk estimates for 2020 using symmetric adaptive smoothing and asymmetric adaptive smoothing with 0.05 high risk tolerance contours denoted by a grey dash dot line.

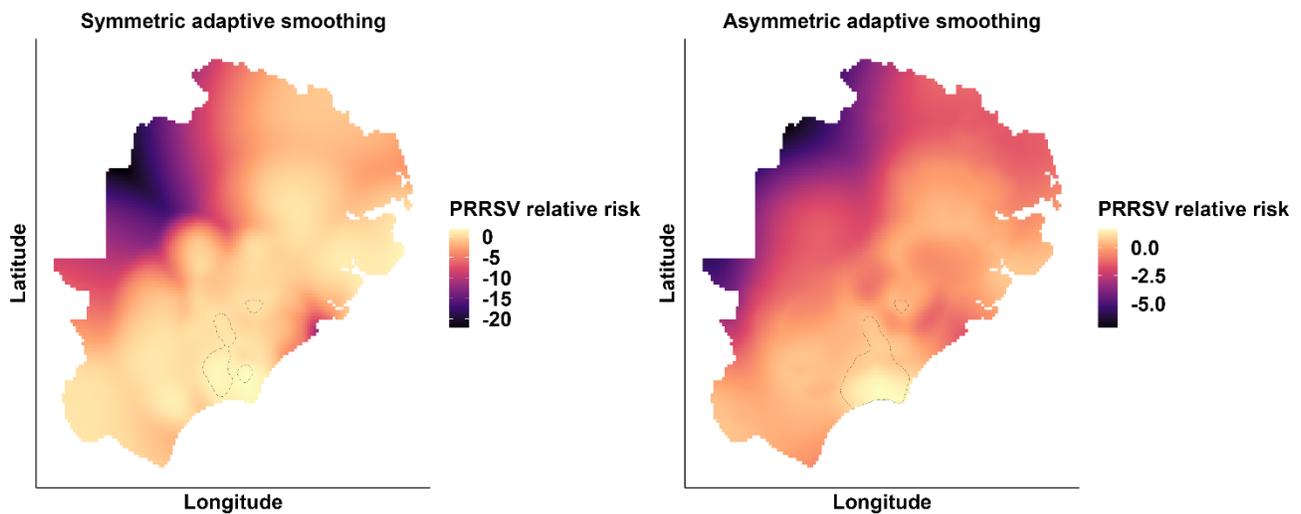

**Supplementary Material Figure S4**. Spatial relative risk estimates for the 2017 - 2018 PRRSV season using symmetric adaptive smoothing and asymmetric adaptive smoothing with 0.05 high risk tolerance contours denoted by a grey dash dot line.

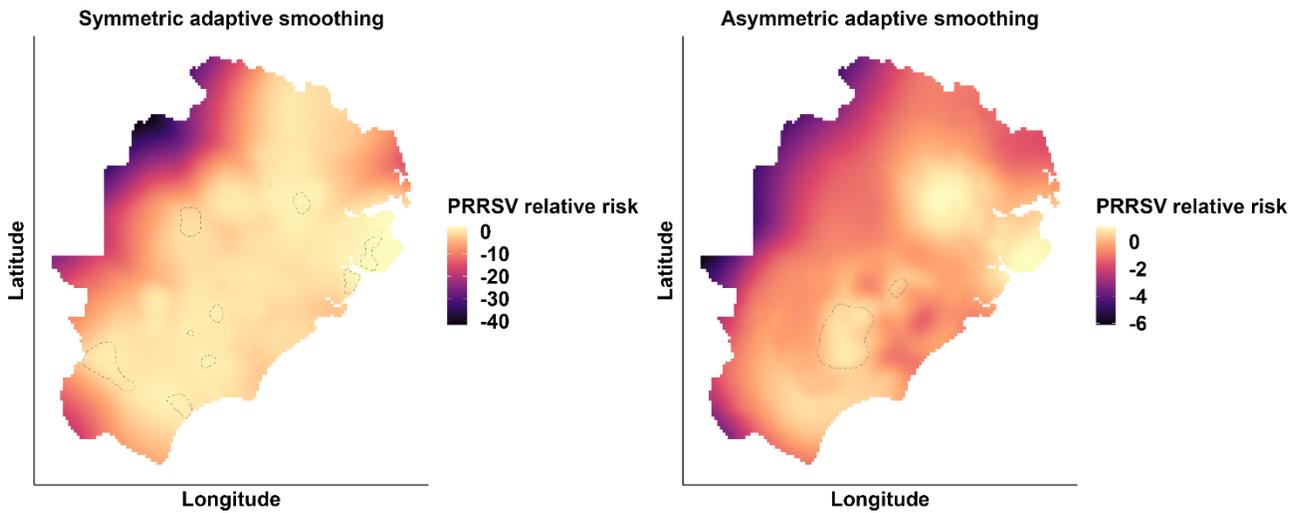

**Supplementary Material Figure S5**. Spatial relative risk estimates for the 2018 - 2019 PRRSV season using symmetric adaptive smoothing and asymmetric adaptive smoothing with 0.05 high risk tolerance contours denoted by a grey dash dot line.

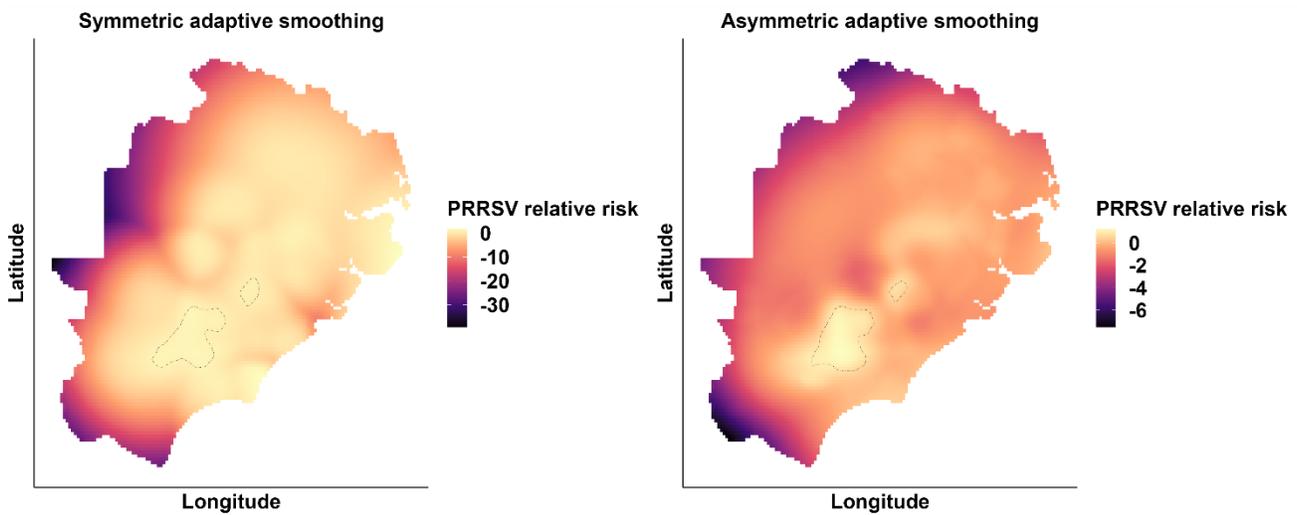

**Supplementary Material Figure S6**. Spatial relative risk estimates for the 2019 - 2020 PRRSV season using symmetric adaptive smoothing and asymmetric adaptive smoothing with 0.05 high risk tolerance contours denoted by a grey dash dot line.

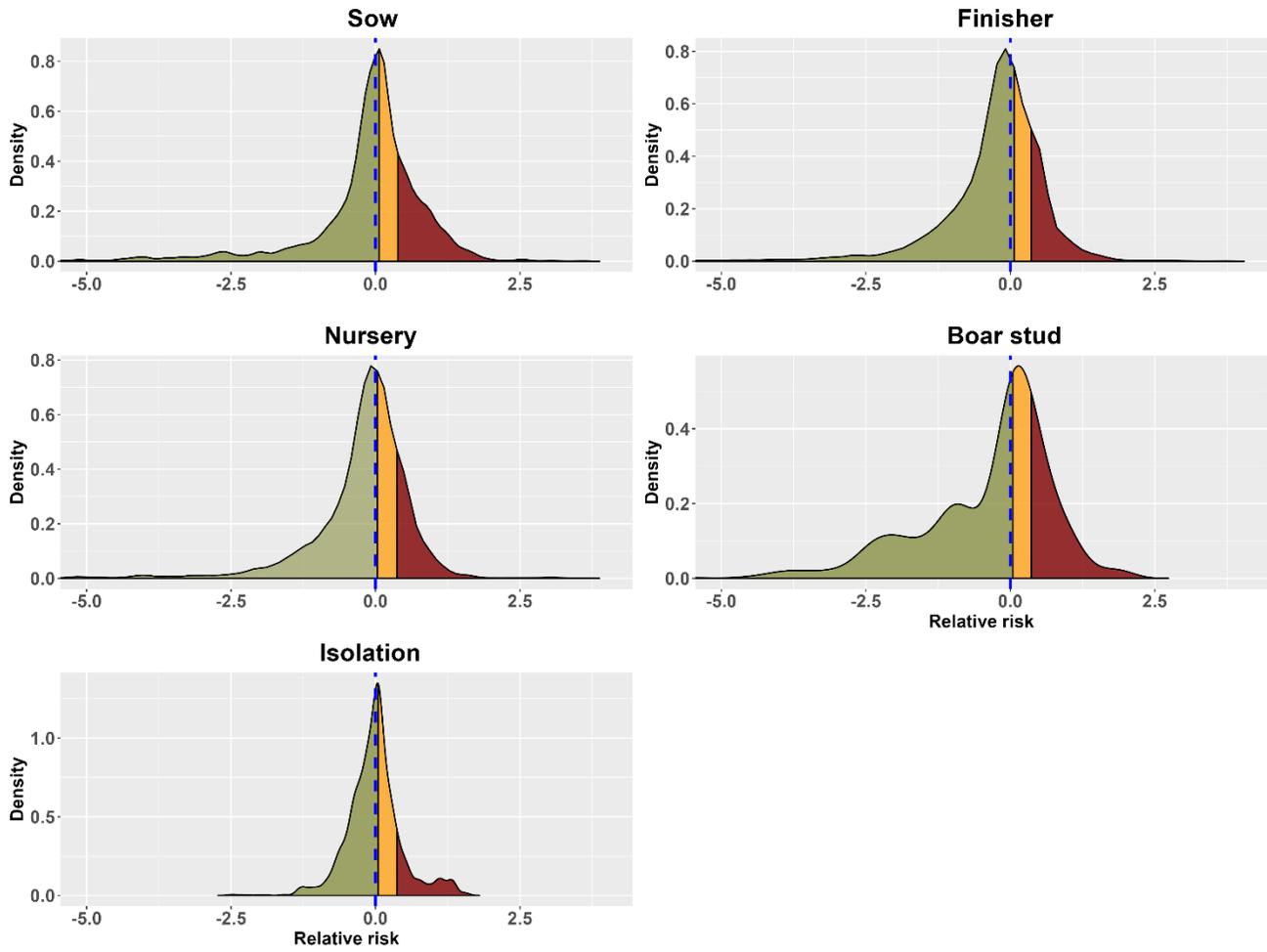

**Supplementary Material Figure S7.** Density plot of the spatiotemporal distribution of relative risk values for the year 2018.

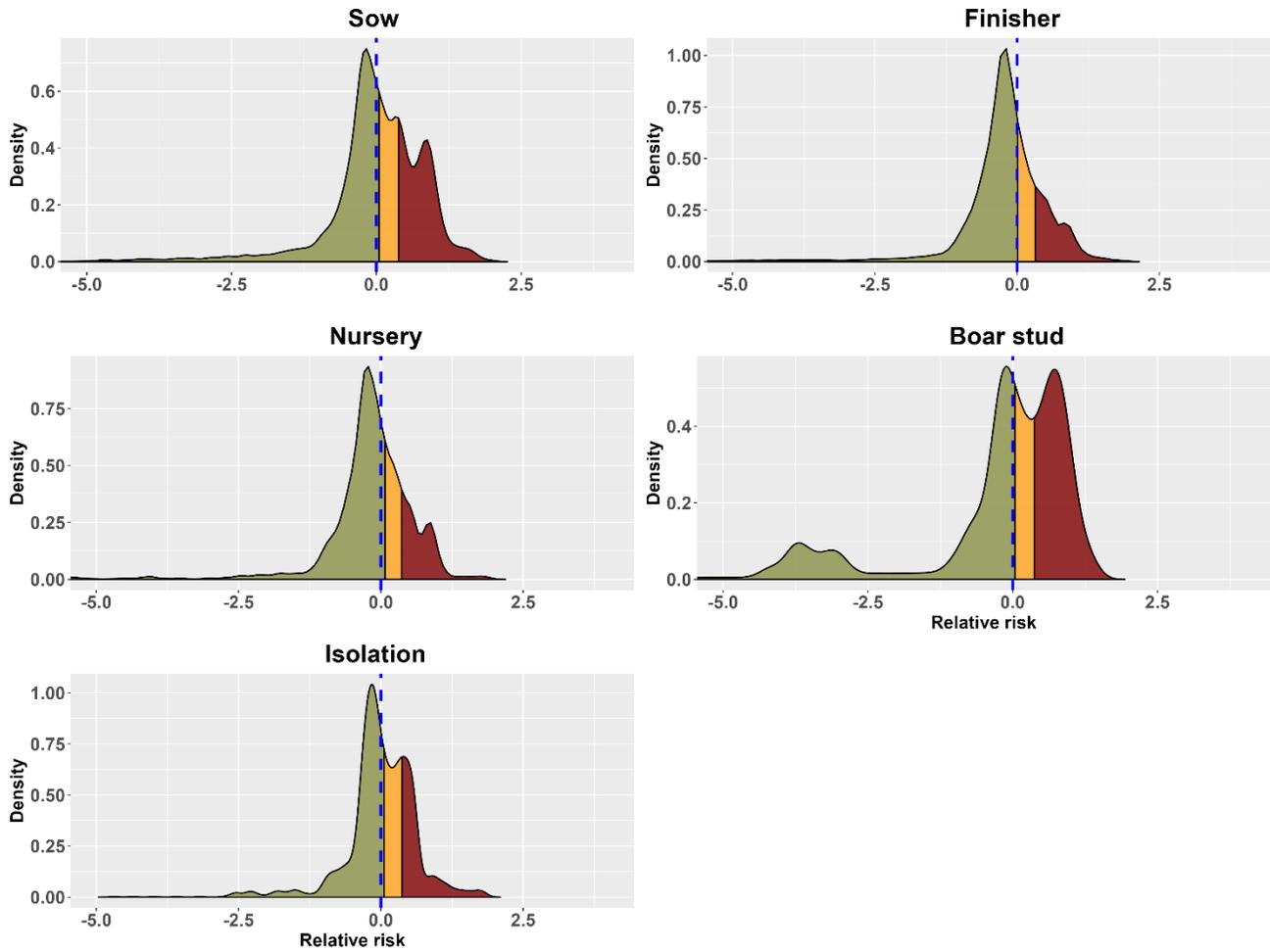

**Supplementary Material Figure S8.** Density plot of the spatiotemporal distribution of relative risk values for the year 2019.

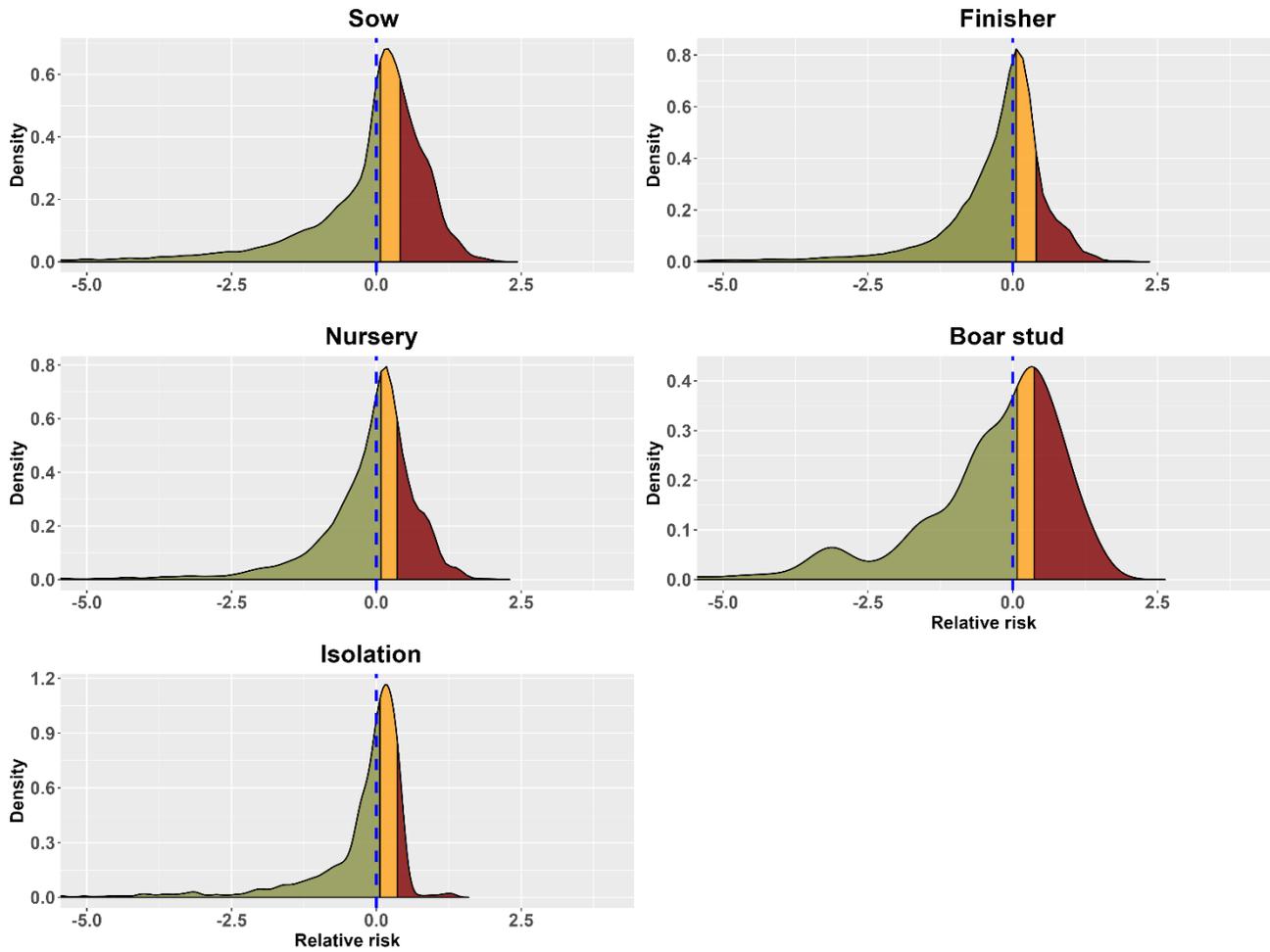

**Supplementary Material Figure S9.** Density plot of the spatiotemporal distribution of relative risk values for the year 2020.

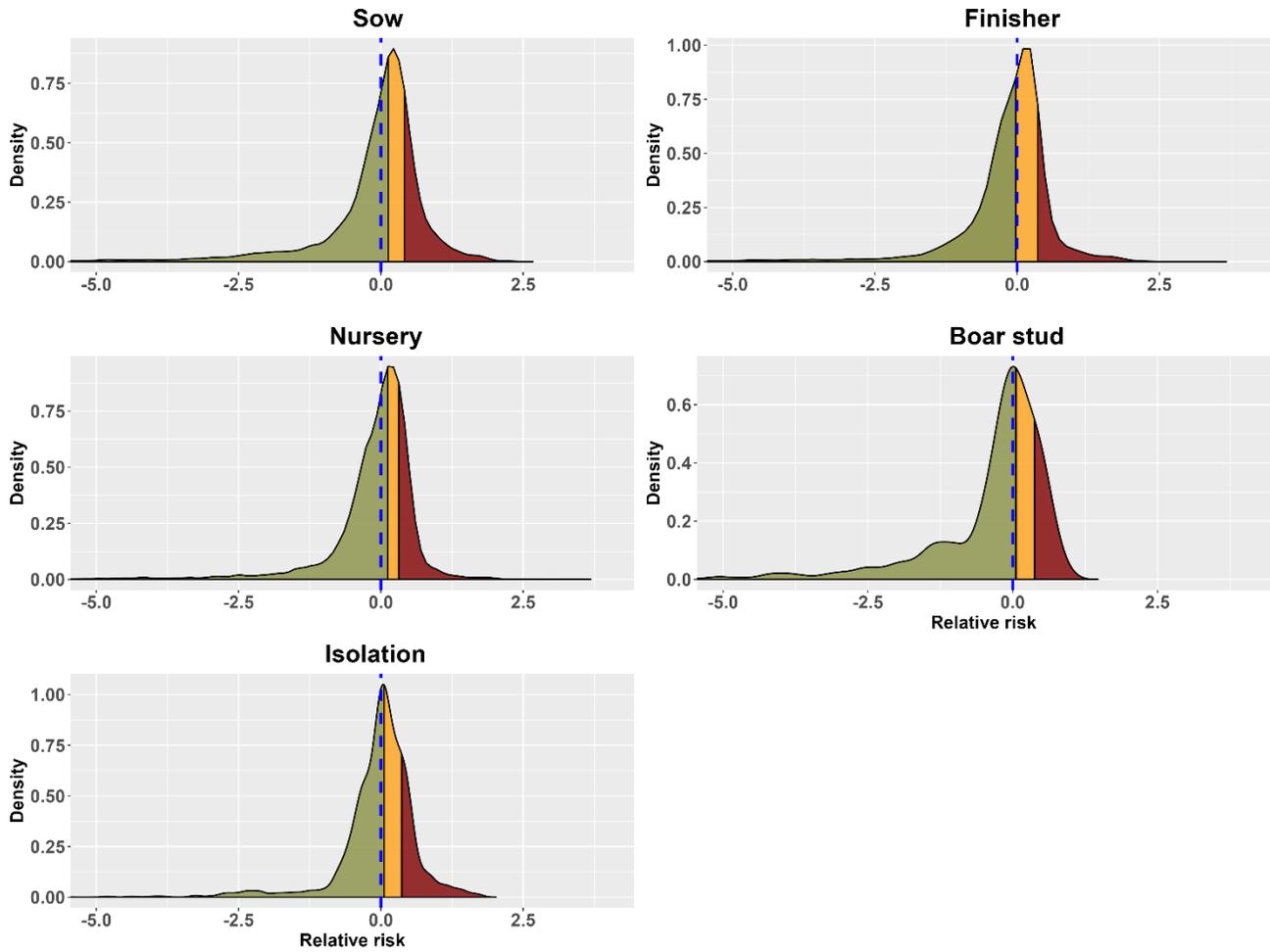

**Supplementary Material Figure S10.** Density plot of the spatiotemporal distribution of relative risk values for the 2017 - 2018 PRRSV season.

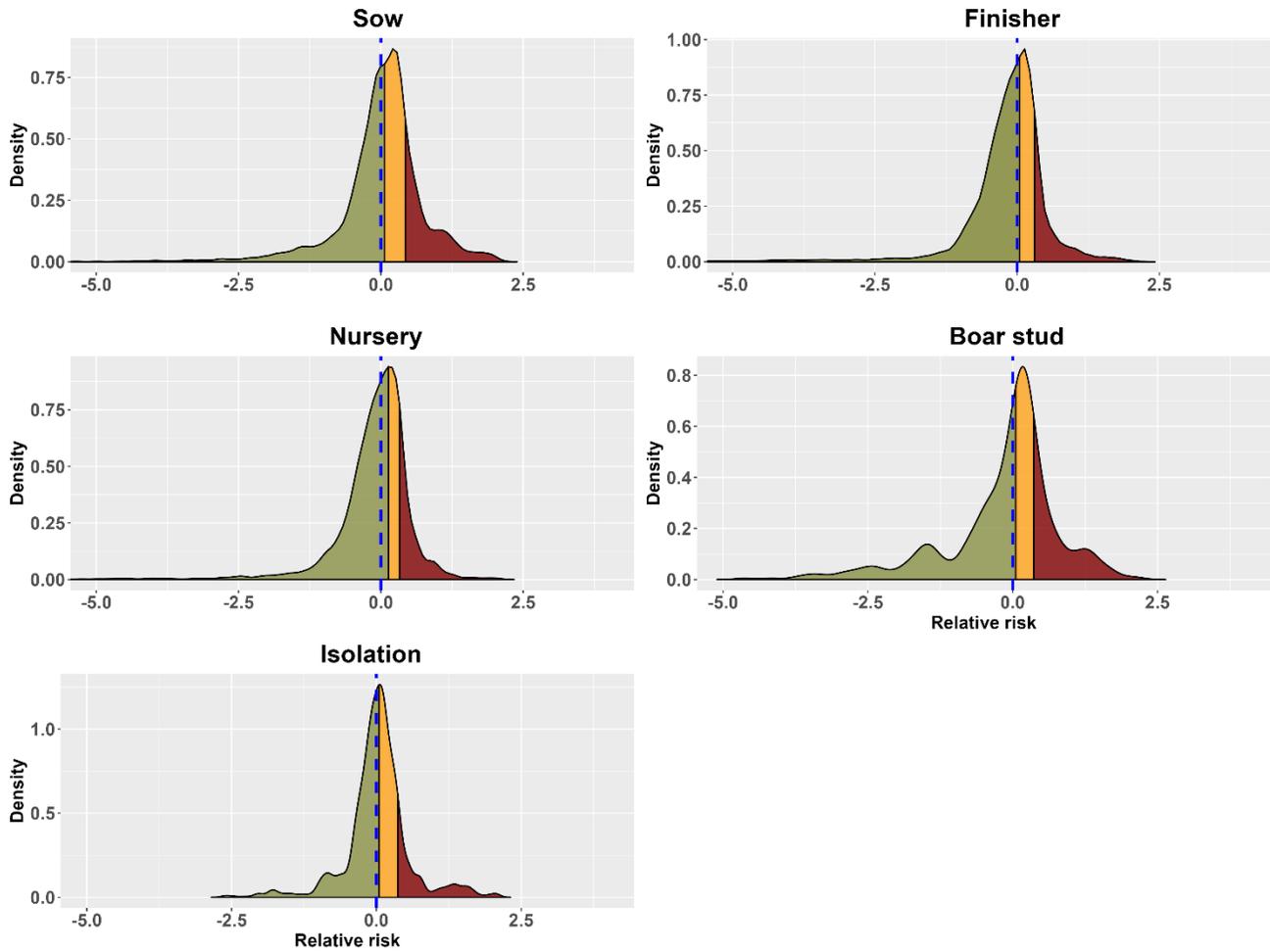

**Supplementary Material Figure S11.** Density plot of the spatiotemporal distribution of relative risk values for the 2018 - 2019 PRRSV season.

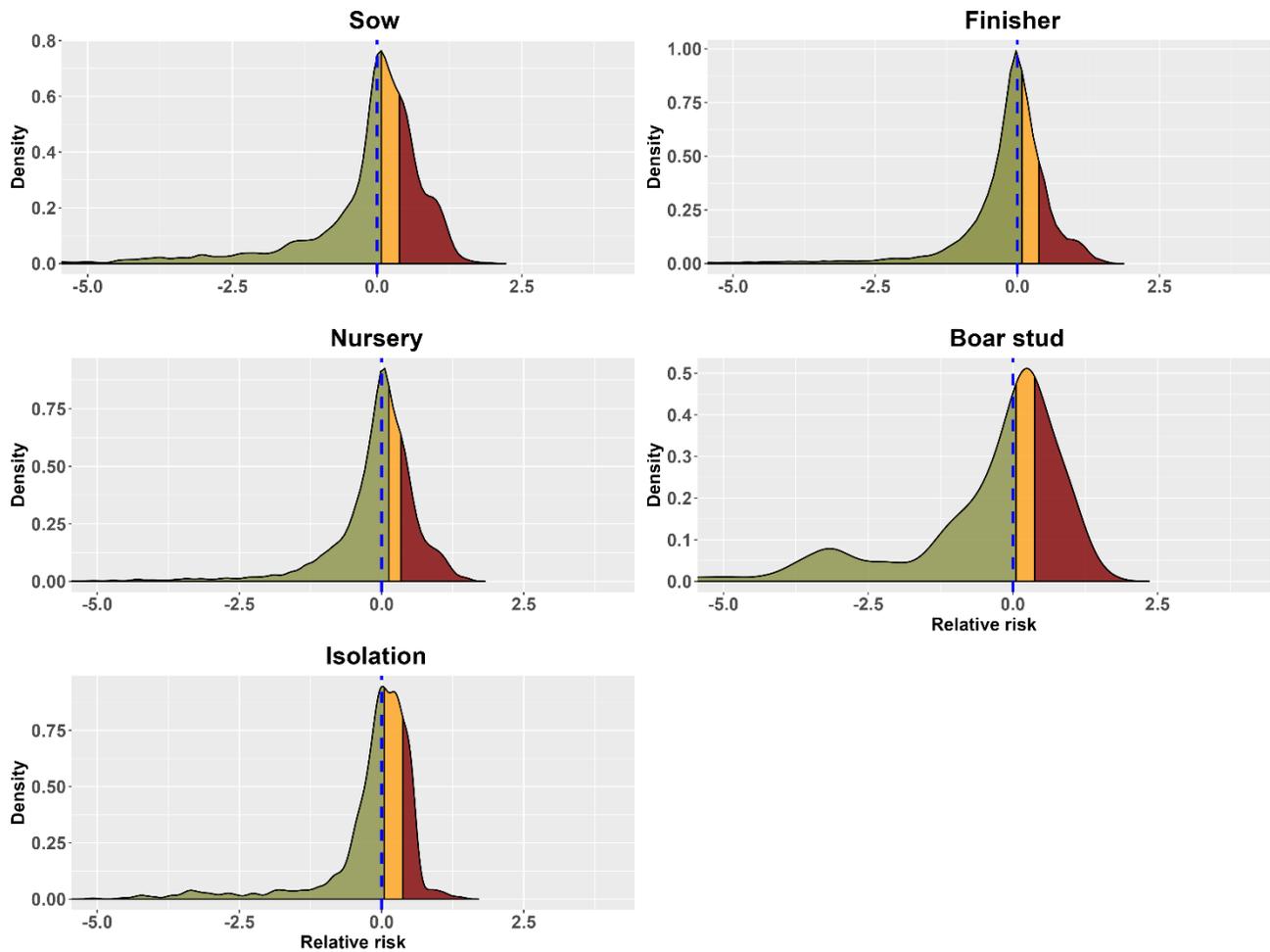

**Supplementary Material Figure S12.** Density plot of the spatiotemporal distribution of relative risk values for the 2019 - 2020 PRRSV season.

**Supplementary Material Table S1.** Description of variables considered in the Bayesian spatiotemporal hierarchical model.

| Parameter | Definition | Reference |
|---|---|---|
| **Enhanced vegetation index (EVI)** | EVI data was downloaded at a 16-day temporal resolution and were used to calculate weekly averages by using one observation every 15 days. EVI was chosen over The Normalized Difference Vegetation Index since it corrects for some atmospheric conditions, and canopy background noise, and is more sensitive in areas with dense vegetation | National Aeronautics and Space Administration (NASA), MODIS Land Products (Huete et al., 2002) |
| **Aboveground biomass density** | Global Ecosystem Dynamics Investigation (GEDI) Level 4 B products offer estimates of aboveground biomass density in megagrams per hectare (Mg/ha) at a 1km x 1km resolution. | Oak Ridge National Laboratory, Distributed Active Archive Center for Biogeochemical Dynamics website (ORNL DAAC, 2022) |
| **Canopy height** | Global Ecosystem Dynamics Investigation (GEDI) Level 3 data provided as an average (meters) of the received waveform signal that was first reflected of the canopy (canopy height) relative to the WGS84 reference ellipsoid. | Oak Ridge National Laboratory, Distributed Active Archive Center for Biogeochemical Dynamics website (ORNL DAAC, 2022) |
| **Land surface elevation** | Global Ecosystem Dynamics Investigation (GEDI) Level 3 data provided as an average (meters) of the lowest received waveform (land surface elevation) signal received relative to the WGS84 reference ellipsoid. | Oak Ridge National Laboratory, Distributed Active Archive Center for Biogeochemical Dynamics website (ORNL DAAC, 2022) |
| **Number of days temperature was between 4°C and 10°C (T[4°C,10°C])** | Daily minimum and maximum land surface temperature was used to calculate the daily average land surface temperature following the methodology described by Spangler et al., 2019. | Daymet: Daily Surface Weather Data (Thornton et al., 2020) |
| **Number of days relative humidity was below %20 (RH < 20%)** | Daily water vapor pressure data were used to calculate daily average relative humidity following the methodology described in Spangler et al., 2019. | Daymet: Daily Surface Weather Data (Thornton et al., 2020) |
| **In-degree** | Number of contacts with direction to a specific farm. | (Wasserman and Faust, 1994) |
| **Out-degree** | Number of contacts originating from a specific farm. | (Wasserman and Faust, 1994) |
| **Closeness centrality** | Closeness centrality measures how many steps are required to access every other vertex from a given node; this measure can be directed at incoming steps and otherwise outgoing steps. | (Freeman, 1978) |

| **PageRank** | Google Page Rank is a link analysis algorithm that produces a ranking of importance of all the municipalities of a network with a range of values between zero and one. | (Brin and Page, 1998) |
|---|---|---|
| **Betweenness** | Describes if the nodes and betweenness of edges are (roughly) defined by the number of geodesics (shortest paths) going through a vertex or an edge. | (Freeman, 1978) |
| **Clustering coefficient** | Measures the degree to which nodes in a network tend to cluster together, with a range of values between zero and one. | (Watts and Strogatz, 1998) |
| **Line of separation access points (LOSAP)** | On-farm biosecurity feature representing locations where people or animals cross to gain access to buildings housing animals, and all other areas where employees and equipment have been completely sanitized. Supplementary Material Figure S17. | Rapid Access Biosecurity app™ (Machado et al., 2023) |
| **Perimeter buffer area access points (PBAAP)** | On-farm biosecurity feature locations where people or animals can gain access to the perimeter buffer areas (PBA), which may encompass the entire site or buildings housing animals and may or may not include feed bins. Equipment and vehicles should be sanitized prior to entry. Supply drop-off area and carcass disposal locations may also be in the PBA. Supplementary Material Figure S17. | Rapid Access Biosecurity app™ (Machado et al., 2023) |
| **Site entry (SE)** | Represents the location of entry into farm. Supplementary Material Figure S17. | Rapid Access Biosecurity app™ (Machado et al., 2023) |
| **Pig capacity** | A farm's pig capacity as reported to MSHMP. | Morrison Swine Health Monitoring Project (MSHMP) (Perez et al., 2019) |

| **Farm density** | Variable created by applying a 17 km buffer around a farm's location and producing a count of farms within the buffer. | Extracted using farm location provided by the Morrison Swine Health Monitoring Project (MSHMP) (Perez et al., 2019) |

### Constrained Refined Delaunay Triangulation Over Study Area

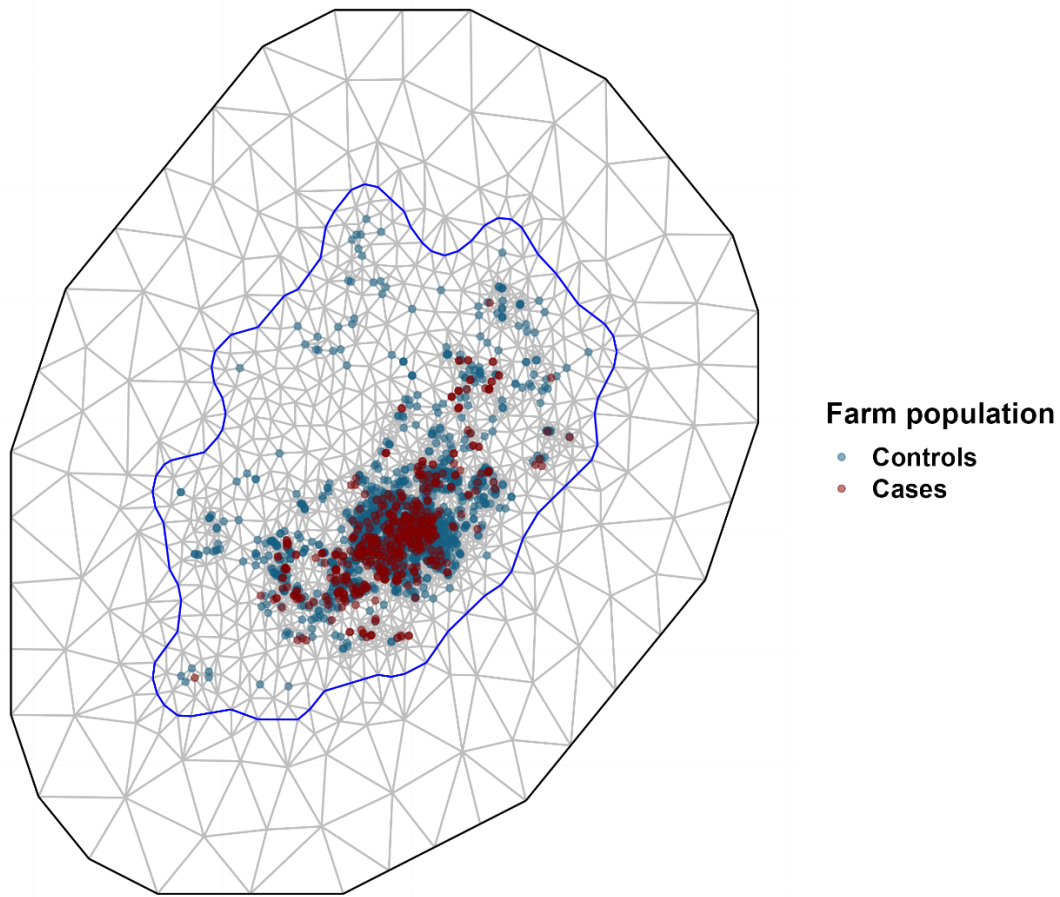

**Farm population**
- Controls
- Cases

**Supplementary Material Figure S13.** Constrained refined Delaunay triangulation over the study area in 2020 with cases (red) and controls (blue) for the entire farm population (n = 2,293).

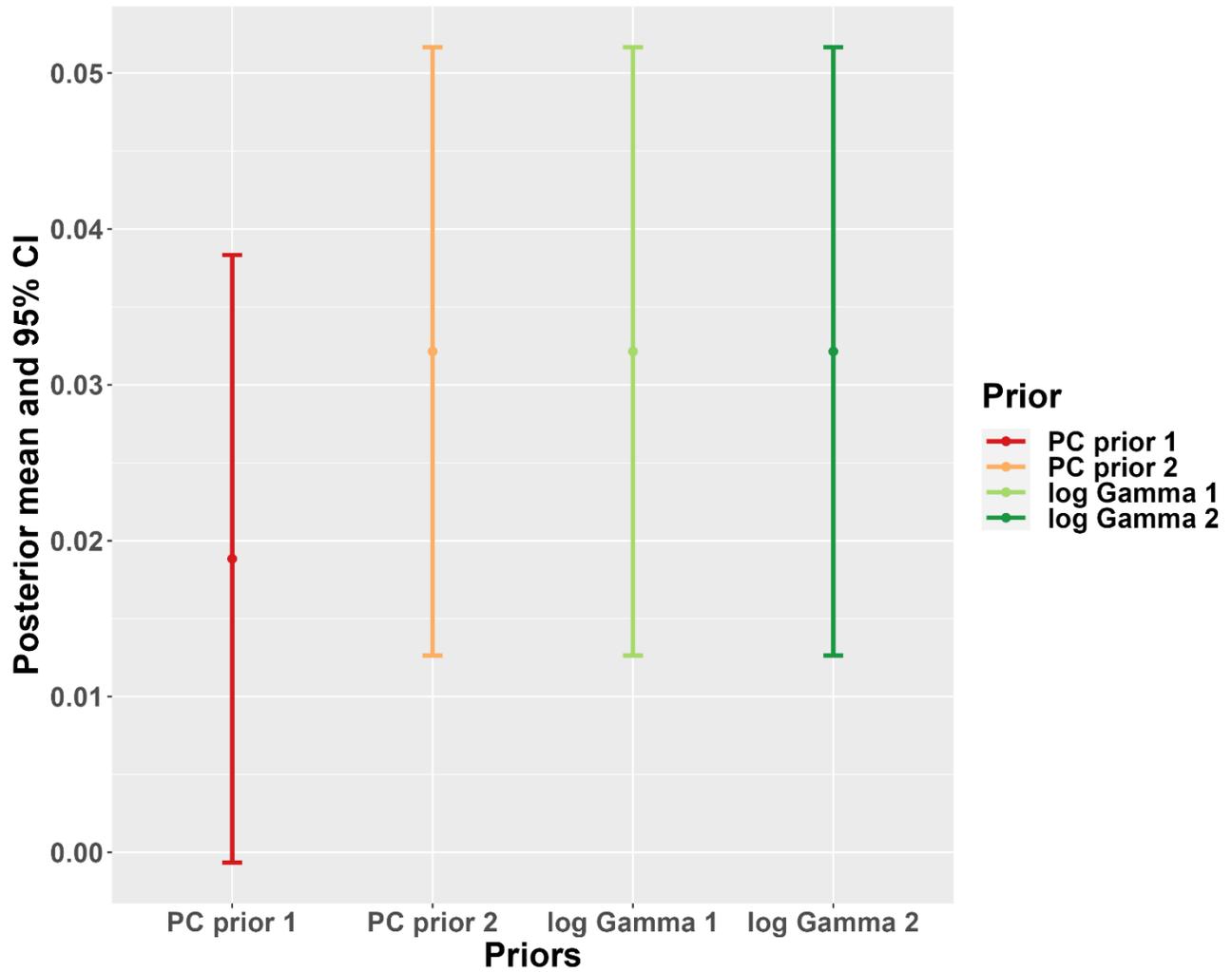

**Supplementary Material Figure S14**. Posterior mean and 95% confidence intervals (CI) for the four priors tested in the sensitivity analysis. All the CI overlap; therefore, the priors do not have a significant impact on the model.

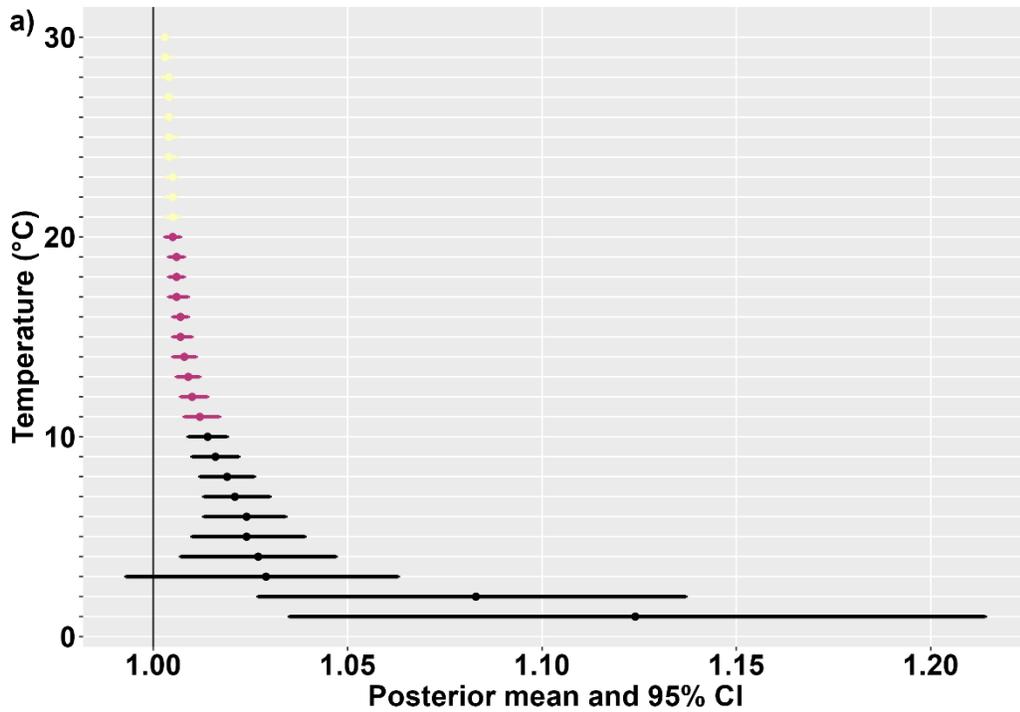

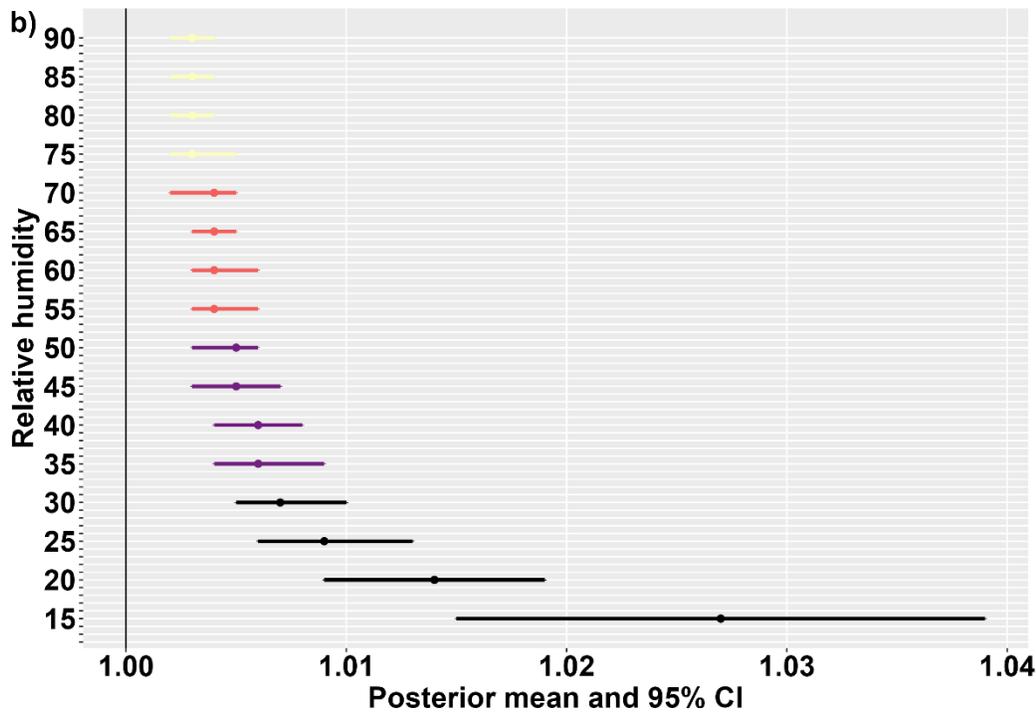

**Supplementary Material Figure S15.** Posterior mean and 95% credible intervals of temperatures (**a**) from 0° C to 30° C and relative humidity (**b**) from 0 to 90%. The Bayesian spatiotemporal hierarchical model was run sequentially with individual temperatures (0° C to 30° C) and individual relative humidities (0 to 90%). Lower temperatures (< 10° C) and relative humidity values (< 20%) represent values with stronger associations with PRRSV outbreaks as opposed to higher temperatures and relative humidity.

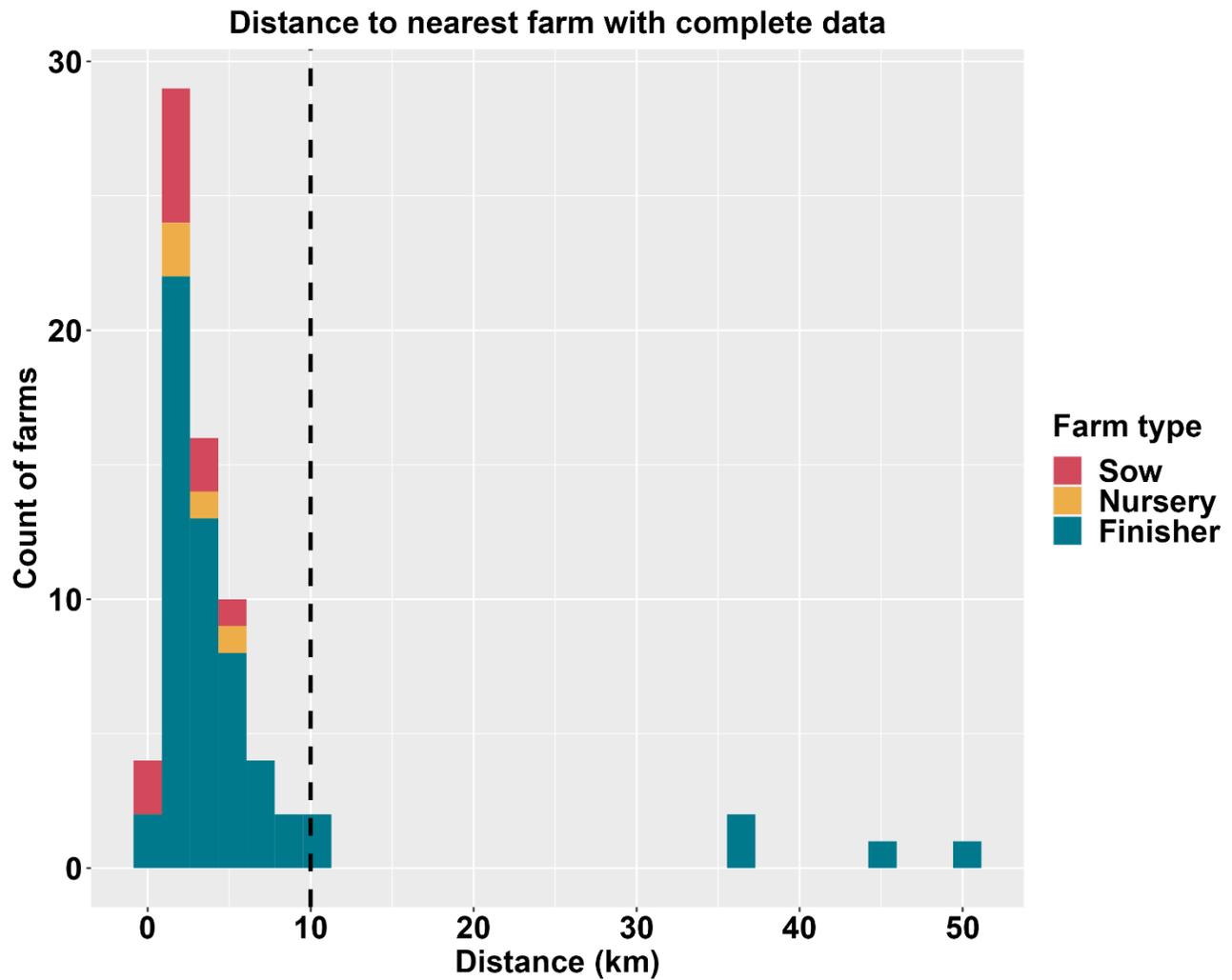

**Supplementary Material Figure S16.** Average distance (km) of the nearest farm with complete data used to fill in missing above-ground biomass density, canopy height, and elevation of 67 out of 71 farms missing data. Four farms exceeded the 10 km cut-off distance. The dashed black line represents the cut-off of acceptable distance to use.

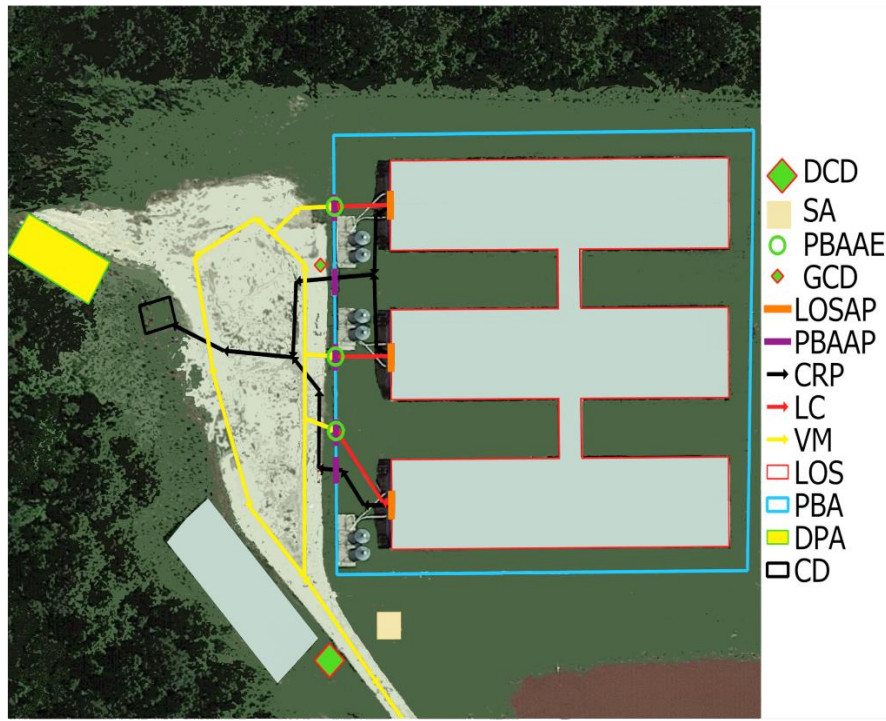

| Legend | Symbol | Legend | Symbol |
|---|---|---|---|
| LOSAE (LOS Animal Emergency) | ◯ | PBAAP (PBA Access Point) | — |
| SE (Site Entry) | ★ | LOSAP (LOS Access Point) | — |
| PBAAE (PBA Access Entry, animals only) | ◯ | CRP (Carcasses Removal Pathways) | → |
| DCD (Designated Cleaning and Disinfection vehicle station) | ◆ | VM (Vehicle Movements) | → |
| GCD (Generalized Cleaning and Disinfection temporary station) | ◆ | LC (Loading Chute) | → |
| PCD (Proposed Cleaning and Disinfection-temporary station) | ✚ | PBA (Perimeter Buffer Area) | ▭ |
| ADU (Dumpster for dead animals) | ◼ | LOS (Line of Separation) | ▭ |
| TD (Trash Dumpster) | ▲ | DPA (Designated Parking Area) | ▭ |
| SA (Supply drop-off Area) | ▣ | CD (Carcasses Disposal location/ADU box) | ▭ |

**Supplementary Material Figure S17.** Map of an example farm within RABapp™ (Machado et al., 2023) showing farm features including Site Entry (blue star), Line of Separation Access Point (LOSAP - orange line), and Perimeter Buffer Area Access Point (PBAAP - purple line).

**Supplementary Material Table S2.** PRRSV season count of cases and controls by farm type within significant ($p < 0.05$) high risk areas estimated using a spatial asymmetric adaptive smoothing approach for 2,293 farms (n = number of farms per farm type) in a dense pig production region of the U.S.

| PRRSV Season | Sow (n = 319) | | Nursery (n = 468) | | Finisher (n = 1458) | | Isolation (n = 33) | | Boar Stud (n = 15) | |
|---|---|---|---|---|---|---|---|---|---|---|
| | Cases in | Controls | Cases | Controls | Cases | Controls | Cases in | Controls | Cases | Controls |

| | Sow | | Nursery | | Finisher | | Isolation | | Boar Stud | |
|---|---|---|---|---|---|---|---|---|---|---|
| **PRRSV Season 2017 - 2018** | 13/319 | 24/319 | 14/468 | 40/468 | 11/1458 | 55/1458 | 0 | 0 | 0 | 0 |
| **PRRSV Season 2018 - 2019** | 28/319 | 45/319 | 5/468 | 69/468 | 6/1458 | 146/1458 | 0 | 7/33 | 0 | 2/15 |
| **PRRSV Season 2019 - 2020** | 35/319 | 35/319 | 6/468 | 73/468 | 9/1458 | 134/1458 | 0 | 7/33 | 0 | 2/15 |

**Supplementary Material Table S3.** Yearly count of cases and controls by farm type within significant ($p < 0.05$) high risk areas estimated using a spatial symmetric adaptive smoothing approach for 2,293 farms (n = number of farms per farm type) in a dense pig production region of the U.S.

| Year | Sow (n = 319) | | Nursery (n = 468) | | Finisher (n = 1458) | | Isolation (n = 33) | | Boar Stud (n = 15) | |
|---|---|---|---|---|---|---|---|---|---|---|
| | Cases | Controls | Cases | Controls | Cases | Controls | Cases | Controls | Cases | Controls |
| **2018** | 45/319 | 56/319 | 14/468 | 86/468 | 26/1,458 | 146/1,458 | 0 | 1/33 | 0 | 3/15 |
| **2019** | 17/319 | 15/319 | 2/468 | 30/468 | 2/1,458 | 25/1,458 | 0 | 0 | 0 | 0 |
| **2020** | 45/319 | 45/319 | 7/468 | 85/468 | 5/1,458 | 127/1,458 | 0 | 4/33 | 0 | 3/15 |

**Supplementary Material Table S4.** PRRSV seasons count of cases and controls by farm type within significant ($p < 0.05$) high risk areas estimated using a spatial symmetric adaptive smoothing approach for 2,293 farms (n = number of farms per farm type) in a dense pig production region of the U.S.

| PRRSV Season | Sow (n = 319) | | Nursery (n = 468) | | Finisher (n = 1458) | | Isolation (n = 33) | | Boar Stud (n = 15) | |
|---|---|---|---|---|---|---|---|---|---|---|
| | Cases | Controls | Cases | Controls | Cases | Controls | Cases | Controls | Cases | Controls |
| **PRRSV Seasson 2017 - 2018** | 18/319 | 29/319 | 15/468 | 48/468 | 10/1,458 | 73/1,458 | 0 | 1/33 | 0 | 0 |
| **PRRSV Season 2018 - 2019** | 7/319 | 13/319 | 1/468 | 27/468 | 2/1,458 | 27/1,458 | 0 | 0 | 0 | 0 |
| **PRRSV Season 2019 - 2020** | 41/319 | 42/319 | 10/468 | 89/468 | 10/1,458 | 168/1,458 | 0 | 10/33 | 0 | 2/15 |

**Supplementary Material Table S5**. PRRSV season median and interquartile range (IQR) proportions of high, medium, and low PRRSV risk levels based on a 60% exceedance risk threshold by farm type. The entire farm population (2,293) is considered each week.

| Year | Sow | | | Nursery | | | Finisher | | | Isolation | | | Boar Stud | | |
|---|---|---|---|---|---|---|---|---|---|---|---|---|---|---|---|
| | High | Med. | Low | High | Med. | Low | High | Med. | Low | High | Med. | Low | High | Med. | Low |
| **PRRSV 2017 - 2018** | 24% (23% - 35%) | 24% (19% - 31%) | 42% (41% - 51%) | 25% (21% - 27%) | 27% (21% - 31%) | 47% (45% - 54%) | 26% (23% - 29%) | 24% (20% - 28%) | 50% (50% - 52%) | 21% (13% - 23%) | 23% (18% - 39%) | 55% (34% - 67%) | 20% (13% - 20%) | 20% (13% - 33%) | 60% (53% - 73%) |
| **PRRSV 2018 - 2019** | 37% (32% - | 15% (12% - | 45% (39% - | 19% (18% - | 15% (11% - | 65% (45% - | 16% (15% - | 15% (8 % - | 69% (49% - | 12% (12% - | 24% (9 % - | 67% (12% - | 33% (27% - | 20% (13% - | 47% (40% - |

| | | | | | | | | | | | | | | |
|---|---|---|---|---|---|---|---|---|---|---|---|---|---|---|
| 39 %) | 26 %) | 50 %) | 21 %) | 33 %) | 71 %) | 17 %) | 34 %) | 76 %) | 17 %) | 58 %) | 85 %) | 40 %) | 27 %) | 53 %) |
| **PRRSV 2019 - 2020** 26 % (25 % - 27 %) | 22 % (18 % - 25 %) | 51 % (48 % - 55 %) | 19 % (16 % - 21 %) | 20 % (17 % - 24 %) | 60 % (56 % - 64 %) | 18 % (15 % - 22 %) | 19 % (16 % - 22 %) | 63 % (56 % - 67 %) | 30 % (24 % - 42 %) | 33 % (25 % - 39 %) | 33 % (27 % - 36 %) | 33 % (27 % - 33 %) | 20 % (13 % - 20 %) | 47 % (47 % - 53 %) |
| **PRRSV Seasons\*** 29 % (24 % - 36 %) | 22 % (15 % - 26 %) | 47 % (41 % - 53 %) | 21 % (18 % - 24 %) | 22 % (15 % - 30 %) | 58 % (47 % - 64 %) | 18 % (16 % - 25 %) | 20 % (14 % - 27 %) | 56 % (50 % - 68 %) | 21 % (12 % - 30 %) | 30 % (17 % - 39 %) | 36 % (27 % - 67 %) | 27 % (20 % - 33 %) | 20 % (13 % - 27 %) | 53 % (47 % - 60 %) |

\* Median and IQR for all PRRSV seasons combined

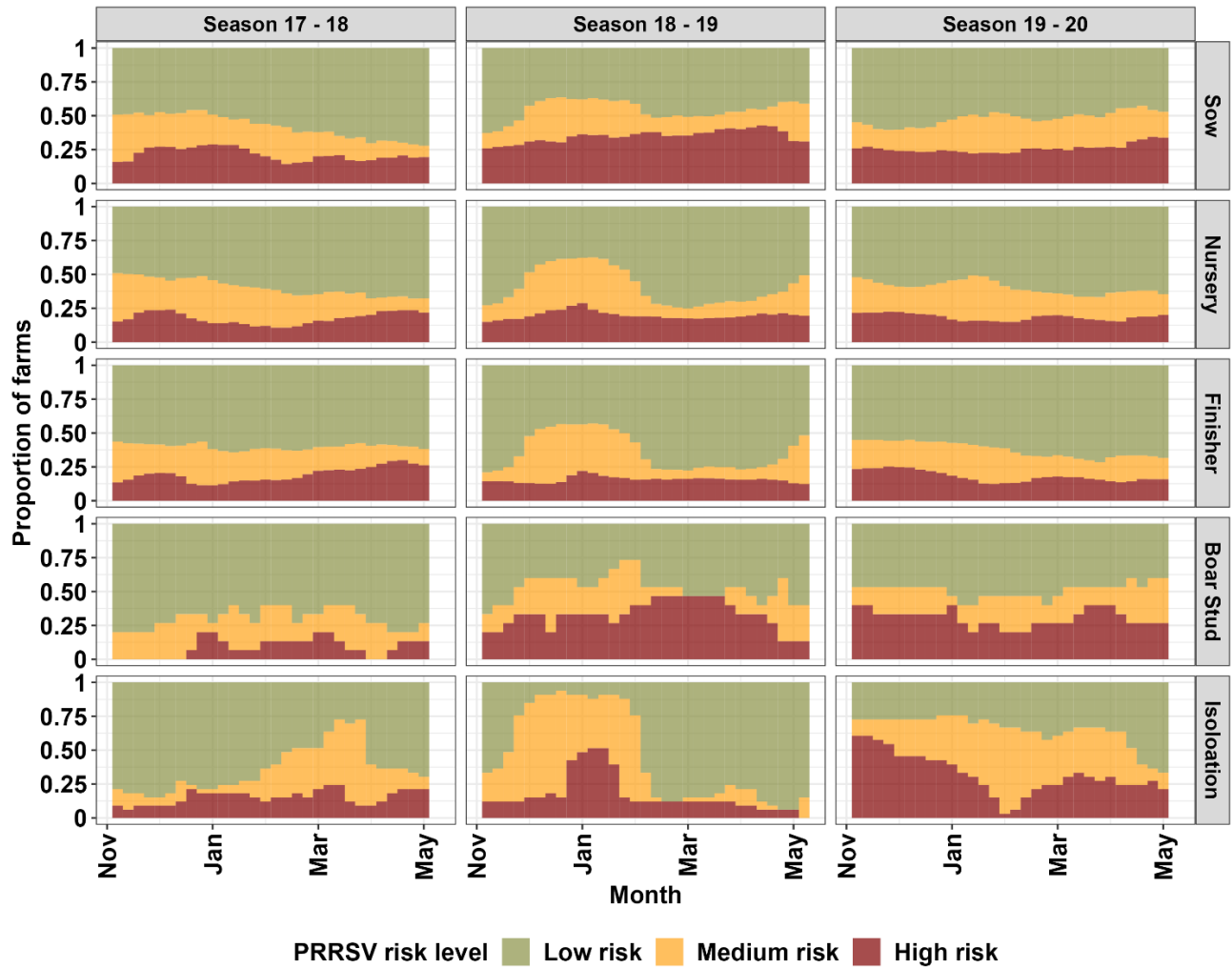

**Supplementary Material Figure S18**. Farm type breakdown of high, medium, and low PRRSV risk levels for the entire farm population (2,293) based on a 60% exceedance risk threshold for a) each week (1 - 30 weeks) in the PRRSV seasonal classifications.

**Supplementary Material Table S6**. Annual percentages of high, medium, and low priority indices (PI) by farm type. PI is calculated as an ordered percentage based on the relative risk value of a farm in reference to the maximum relative risk value of the entire farm population. High, medium, and low categories are then classified by quantile classifications.

| Year | Sow | | | Nursery | | | Finisher | | | Isolation | | | Boar Stud | | |
|---|---|---|---|---|---|---|---|---|---|---|---|---|---|---|---|
| | High | Medium | Low | High | Medium | Low | High | Medium | Low | High | Medium | Low | High | Medium | Low |
| **2018** | 7% | 20 % | 73% | 3% | 14 % | 83% | 4% | 16 % | 80% | 7% | 15% | 78% | 5% | 16 % | 79% |

| **2019** | 14 % | 32 % | 54 % | 5 % | 28 % | 67 % | 7 % | 26 % | 66 % | 8 % | 27% | 65 % | 25 % | 37 % | 39 % |
| **2020** | 12 % | 31 % | 57 % | 7 % | 27 % | 66 % | 8 % | 21 % | 72 % | 1 % | 13 % | 86 % | 14 % | 34 % | 52 % |

**Supplementary Material Table S7**. PRRSV season percentages of high, medium, and low priority indices (PI) by farm type. PI is calculated as an ordered percentage based on the relative risk value of a farm in reference to the maximum relative risk value of the entire farm population. High, medium, and low categories are then classified by quantile classifications.

| Season | Sow | | | Nursery | | | Finisher | | | Isolation | | | Boar stud | | |
|---|---|---|---|---|---|---|---|---|---|---|---|---|---|---|---|
| | High | Medium | Low | High | Medium | Low | High | Medium | Low | High | Medium | Low | High | Medium | Low |
| **PRRSV 2017-2018** | 1 % | 5 % | 94 % | 0.2 % | 4.8 % | 95 % | 2 % | 7 % | 91 % | 0.6 % | 4.4 % | 95 % | 0 | 7 % | 93 % |
| **PRRSV 2018-2019** | 9 % | 29 % | 62 % | 6 % | 31 % | 63 % | 8 % | 34 % | 58 % | 13 % | 36 % | 51 % | 17 % | 31 % | 52 % |
| **PRRSV 2019-2020** | 3 % | 13 % | 84 % | 1 % | 8 % | 91 % | 2 % | 10 % | 88 % | 1 % | 8 % | 91 % | 1 %` | 16 % | 83 % |

**S2. Network analysis**

A total of 55,623 pig movements were used to construct a directed static network of pig movements for the year 2020 and was comprised of 2,286 vertices and 9,111 edges for a total of 104,624,297 pigs moved. An average of 1,881 pigs were moved per shipment.

**Supplementary Material Table S8**. Median and interquartile range (IQR) of node-level metrics calculated for the static pig network by farm type and cases and controls.

| Year | Sow | | Finisher | | Nursery | | Isolation | | Boar Stud | |
|---|---|---|---|---|---|---|---|---|---|---|
| | Cases | Controls | Cases | Controls | Cases | Controls | Cases | Controls | Cases | Controls |
| **Degree** | 8.5 (5 - 13) | 9.5 (6 - 16) | 6 (3 - 12) | 3 (2 - 5) | 11 (8 - 14) | 10 (7 - 14) | - | 5 (4.2 - 6.8) | - | 2 (2 - 3) |
| **In-degree** | 3 (1 - 4) | 2 (1 - 4) | 4 (3 - 11) | 3 (2 - 4) | 2 (1 - 4) | 2 (1 - 3) | - | 5 (2 - 5) | - | 2 (2 - 3) |
| **Out-degree** | 6 (4 - 10) | 6 (3 - 10) | 0 (0 - 1) | 0 (0 - 1) | 9 (7 - 11) | 8 (6 - 11) | - | 1 (0 - 1) | - | 0 (0) |
| **Closeness centrality** | 0.0000847 (0.0000633 - 0.0000910) | 0.0000856 (0.0000788 - 0.0000907) | 0.0000692 (0.0000649 - 0.0000930) | 0.0000727 (0.0000649 - 0.0000804) | 0.0000774 (0.0000718 - 0.0000837) | 0.0000776 (0.0000718 - 0.0000841) | - | 0.0000575 (0.0000574 - 0.0000575) | - | 0.0000678 (0.0000678 - 0.0000678) |
| **Betweenness** | 361 (118 - 731) | 318 (107 - 804) | 0 (0 - 2) | 0 (0) | 186 (120 - 340) | 128 (53 - 241) | - | 0 (0 - 126) | - | 0 (0) |
| **Clustering Coefficient** | 0 (0 - 0.018) | 0 (0 - 0.04) | 0 (0 - 0.01) | 0 (0) | 0 (0 - 0.018) | 0 (0 - 0.015) | - | 0 (0 - 0.003) | - | 0 (0 - 0.17) |
| **Page Rank** | 0.000309 (0.000272 - 0.000403) | 0.000304 (0.000268 - 0.000359) | 0.000365 (0.000321 - 0.000517) | 0.000329 (0.000292 - 0.000378) | 0.000333 (0.000310 - 0.000420) | 0.000319 (0.000281 - 0.000377) | | 0.000292 (0.000261 - 0.000298) | - | 0.000265 (0.000258 - 0.000305) |